\pdfoutput=1
\documentclass[twocolumn]{aastex63}

\usepackage{siunitx}
\DeclareSIUnit\mEarth{M_\oplus}
\DeclareSIUnit\mSun{M_\odot}
\DeclareSIUnit\rEarth{R_\oplus}
\DeclareSIUnit\year{yr}
\DeclareSIUnit\au{au}
\DeclareSIUnit\dex{dex}
\DeclareSIUnit\ppm{ppm}

\usepackage{lipsum}
\usepackage{multirow}
\usepackage{comment}						 %
\includecomment{comment}
\specialcomment{note}{\begingroup\sffamily\color{gray}}{\endgroup}

\usepackage[disable]{todonotes}			%

\received{June 1, 2999}
\revised{January 10, 2999}
\accepted{October 3, 2020}

\shorttitle{TIC 237913194b}
\shortauthors{Schlecker et al.}
\graphicspath{{./}{fig/}}

\begin{document}

\title{A Highly Eccentric Warm Jupiter Orbiting TIC 237913194}

\newcommand{\feh}{\ensuremath{{\rm [Fe/H]}}}
\newcommand{\teff}{\ensuremath{T_{\rm eff}}}
\newcommand{\teq}{\ensuremath{T_{\rm eq}}}
\newcommand{\logg}{\ensuremath{\log{g}}}
\newcommand{\zaspe}{\texttt{ZASPE}}
\newcommand{\ceres}{\texttt{CERES}}
\newcommand{\tess}{\textit{TESS}}
\newcommand{\chat}{\textit{CHAT}}
\newcommand{\lcogt}{\textit{LCOGT}}
\newcommand{\vsini}{\ensuremath{v \sin{i}}}
\newcommand{\kms}{\ensuremath{{\rm km\,s^{-1}}}}
\newcommand{\mjup}{\ensuremath{{\rm M_{J}}}}
\newcommand{\mearth}{\ensuremath{{\rm M}_{\oplus}}}
\newcommand{\mpl}{\ensuremath{{\rm M_P}}}
\newcommand{\rjup}{\ensuremath{{\rm R_J}}}
\newcommand{\rpl}{\ensuremath{{R_\mathrm{P}}}}
\newcommand{\rstar}{\ensuremath{{\rm R}_{\star}}}
\newcommand{\mstar}{\ensuremath{{\rm M}_{\star}}}
\newcommand{\lstar}{\ensuremath{{\rm L}_{\star}}}
\newcommand{\rsun}{\ensuremath{{\rm R}_{\odot}}}
\newcommand{\msun}{\ensuremath{{\rm M}_{\odot}}}
\newcommand{\lsun}{\ensuremath{{\rm L}_{\odot}}}

\newcommand{\mptess}{\ensuremath{1.942_{-0.091}^{+0.091}}}
\newcommand{\rptess}{\ensuremath{1.117_{-0.047}^{+0.054}}}
\newcommand{\mst}{\ensuremath{1.026_{-0.055}^{+0.057}}}
\newcommand{\rst}{\ensuremath{1.088_{-0.012}^{+0.012}}}
\newcommand{\age}{\ensuremath{5.7 \pm 1.7 }}
\newcommand{\teqv}{\ensuremath{974}}

\newcommand{\period}{\ensuremath{15.168865 \pm 0.000018 }}
\newcommand{\sma}{\ensuremath{0.1207_{-0.0037}^{+0.0037} }}
\newcommand{\rhost}{\ensuremath{1.12 \pm 0.11 }}

\newcommand{\TC}{\ensuremath{2458319.15055^{+0.00077}_{-0.00077}}}
\newcommand{\imp}{\ensuremath{0.900^{+0.017}_{-0.017}}}

\newcommand{\teffv}{\ensuremath{5788 \pm 80}}
\newcommand{\loggv}{\ensuremath{4.376 \pm 0.021}}
\newcommand{\fehv}{\ensuremath{+0.14 \pm 0.05}}
\newcommand{\vsiniv}{\ensuremath{2.18 \pm 0.41}}

\newcommand{\plname}{TIC~237913194b}
\newcommand{\stname}{TIC~237913194}
\newcommand{\rhopl}{\ensuremath{{\rm \rho_P}}}
\newcommand{\gccm}{\ensuremath{\mathrm{g}\,\mathrm{cm}^{-3}}}

\newcommand{\rev}[1]{{#1}}

\correspondingauthor{Martin Schlecker}
\email{schlecker@mpia.de}

\author[0000-0001-8355-2107]{Martin Schlecker} %
\affiliation{Max-Planck-Institut f\"ur Astronomie, K\"onigstuhl 17, Heidelberg 69117, Germany}

\author[0000-0002-0436-7833]{Diana Kossakowski}
\affiliation{Max-Planck-Institut f\"ur Astronomie, K\"onigstuhl 17, Heidelberg 69117, Germany}

\author[0000-0002-9158-7315]{Rafael Brahm}
\affiliation{Facultad de Ingeniería y Ciencias, Universidad Adolfo Ib\'a\~nez, Av.\ Diagonal las Torres 2640, Pe\~nalol\'en, Santiago, Chile}
\affiliation{Millennium Institute for Astrophysics, Chile}

\author[0000-0001-9513-1449]{N\'{e}stor Espinoza}
\affiliation{Space Telescope Science Institute, 3700 San Martin Drive, Baltimore, MD 21218, USA}

\author[0000-0002-1493-300X]{Thomas Henning}
\affiliation{Max-Planck-Institut f\"ur Astronomie, K\"onigstuhl 17, Heidelberg 69117, Germany}

\author[0000-0001-9355-3752]{Ludmila Carone}
\affiliation{Max-Planck-Institut f\"ur Astronomie, K\"onigstuhl 17, Heidelberg 69117, Germany}

\author[0000-0002-0502-0428]{Karan Molaverdikhani}
\affiliation{\rev{Landessternwarte, Zentrum für Astronomie der Universität Heidelberg, Königstuhl 12, 69117 Heidelberg, Germany}}
\affiliation{Max-Planck-Institut f\"ur Astronomie, K\"onigstuhl 17, Heidelberg 69117, Germany}

\author[0000-0002-0236-775X]{Trifon Trifonov}
\affiliation{Max-Planck-Institut f\"ur Astronomie, K\"onigstuhl 17, Heidelberg 69117, Germany}

\author[0000-0003-4096-7067]{Paul Mollière}
\affiliation{Max-Planck-Institut f\"ur Astronomie, K\"onigstuhl 17, Heidelberg 69117, Germany}

\author[0000-0002-5945-7975]{Melissa J. Hobson}
\affiliation{Millennium Institute for Astrophysics, Chile}

\author[0000-0002-5389-3944]{Andr\'es Jord\'an} %
\affiliation{Facultad de Ingeniería y Ciencias, Universidad Adolfo Ib\'a\~nez, Av.\ Diagonal las Torres 2640, Pe\~nalol\'en, Santiago, Chile}
\affiliation{Millennium Institute for Astrophysics, Chile}

\author{Felipe I. Rojas}
\affiliation{Instituto de Astrof\'isica, Pontificia Universidad Cat\'olica de Chile, Av.\ Vicu\~na Mackenna 4860, Macul, Santiago, Chile}
\affiliation{Millennium Institute for Astrophysics, Chile}

    \author[0000-0002-8227-5467]{Hubert Klahr}
    \affiliation{Max-Planck-Institut f\"ur Astronomie, K\"onigstuhl 17, Heidelberg 69117, Germany}

\author[0000-0001-8128-3126]{Paula Sarkis}
\affiliation{Max-Planck-Institut f\"ur Astronomie, K\"onigstuhl 17, Heidelberg 69117, Germany}

\author[0000-0001-7204-6727]{G\'asp\'ar \'A.\ Bakos} %
\affiliation{Department of Astrophysical Sciences, Princeton University, NJ 08544, USA}

\author[0000-0002-0628-0088]{Waqas Bhatti} %
\affiliation{Department of Astrophysical Sciences, Princeton University, NJ 08544, USA}

\author{David Osip} %
\affiliation{Las Campanas Observatory, Carnegie Institution of Washington, Colina el Pino, Casilla 601 La Serena, Chile}

\author[0000-0001-7070-3842]{Vincent Suc} %
\affiliation{Facultad de Ingeniería y Ciencias, Universidad Adolfo Ib\'a\~nez, Av.\ Diagonal las Torres 2640, Pe\~nalol\'en, Santiago, Chile}
\affiliation{El Sauce Observatory, Chile}

\author{George Ricker} %
\affiliation{Department of Physics and Kavli Institute for Astrophysics and Space Research, Massachusetts Institute of Technology, Cambridge, MA 02139, USA}

\author{Roland Vanderspek} %
\affiliation{Department of Physics and Kavli Institute for Astrophysics and Space Research, Massachusetts Institute of Technology, Cambridge, MA 02139, USA}

\author{David W. Latham}%
\affiliation{Center for Astrophysics \textbar \ Harvard \& Smithsonian, 60 Garden St, Cambridge, MA 02138, USA}

\author[0000-0002-6892-6948]{Sara Seager} %
\affiliation{Department of Physics and Kavli Institute for Astrophysics and Space Research, Massachusetts Institute of Technology, Cambridge, MA 02139, USA}
\affiliation{Department of Earth, Atmospheric and Planetary Sciences, Massachusetts Institute of Technology, Cambridge, MA 02139, USA}
\affiliation{Department of Aeronautics and Astronautics, MIT, 77 Massachusetts Avenue, Cambridge, MA 02139, USA}

\author{Joshua N. Winn} %
\affiliation{Department of Astrophysical Sciences, Princeton University, NJ 08544, USA}

\author[0000-0002-4715-9460]{Jon M. Jenkins} %
\affiliation{NASA Ames Research Center, Moffett Field, CA 94035, USA}

\author{Michael~Vezie} %
\affiliation{Department of Physics and Kavli Institute for Astrophysics and Space Research, Massachusetts Institute of Technology, Cambridge, MA 02139, USA}

\author{Jesus Noel Villase{\~n}or} %
\affiliation{Department of Physics and Kavli Institute for Astrophysics and Space Research, Massachusetts Institute of Technology, Cambridge, MA 02139, USA}

\author[0000-0003-4724-745X]{Mark E. Rose} %
\affiliation{NASA Ames Research Center, Moffett Field, CA 94035, USA}

\author[0000-0003-1286-5231]{David R. Rodriguez} %
\affiliation{Space Telescope Science Institute, 3700 San Martin Drive, Baltimore, MD 21218, USA}

\author[0000-0001-8812-0565]{Joseph E. Rodriguez} %
\affiliation{Center for Astrophysics \textbar \ Harvard \& Smithsonian, 60 Garden St, Cambridge, MA 02138, USA}

\author[0000-0002-8964-8377]{Samuel N. Quinn} %
\affiliation{Center for Astrophysics \textbar \ Harvard \& Smithsonian, 60 Garden St, Cambridge, MA 02138, USA}

\author[0000-0002-1836-3120]{Avi Shporer} %
\affiliation{Department of Physics and Kavli Institute for Astrophysics and Space Research, Massachusetts Institute of Technology, Cambridge, MA 02139, USA}

\begin{abstract}
The orbital parameters of warm Jupiters serve as a record of their formation history, providing constraints on formation scenarios for giant planets on close and intermediate orbits.
Here, we report the discovery of \plname, detected in full frame images from Sectors 1 and 2 of \tess, ground-based photometry (\textit{CHAT}, \textit{LCOGT}), and \textit{FEROS} radial velocity time series.
We constrain its mass to \mpl = \mptess\ \mjup \, and its radius to \rpl = \rptess\ \rjup, implying a bulk density similar to Neptune's.
It orbits a G-type star (\mstar = \mst\ \msun, $V = 12.1$ mag) with a period of \SI{15.17}{\day} on one of the most eccentric orbits of all known warm giants \rev{($e \approx 0.58$)}.
This extreme dynamical state points to a past interaction with an additional, undetected massive companion.
A tidal evolution analysis showed a large tidal dissipation timescale, suggesting that the planet is not a progenitor for a hot Jupiter caught during its high-eccentricity migration.
\plname\ further represents an attractive opportunity to study the energy deposition and redistribution in the atmosphere of a warm Jupiter with high eccentricity.
\end{abstract}


\section{Introduction} \label{sec:intro}
Gravitational interactions among massive planets during their formation and evolution leave an imprint on their orbital parameters.
However, these imprints are frequently erased in the case of hot Jupiters, which are prone to orbital changes through tidal interactions with their host star \citep[e.g.,][]{Eggleton1998}.
Planets on more distant orbits ($P\gtrsim\SI{10}{\day}$), although not as readily detected, are expected to retain this information and thereby provide valuable insights into the formation history of their planetary system.
Unfortunately, the sample of confirmed, nearby transiting warm Jupiters is still small.
The transit survey currently performed by the Transiting Exoplanet Survey Satellite \citep[\tess,][]{Ricker2014} is changing that: hundreds of giant planets on intermediate orbits are expected to be detected during the all-sky survey~\citep{Sullivan2015,Barclay2018a}.
With this in mind, the Warm gIaNts with tEss \citep[WINE,][]{hd1397,jordan:2020} collaboration embarked on a search for such warm Jupiters.
Using a network of photometric and spectroscopic facilities, we identify and follow up \tess\ planet candidates to confirm them, characterize their orbital parameters, and use them to inform planet formation theory.

Here, we report the discovery of a temperate giant planet in a highly eccentric orbit around a G3~star.
By the aid of additional ground-based photometry from the \textit{CHAT} and \textit{LCOGT} telescopes, as well as precise radial velocity measurements from \textit{FEROS}, we were able to tightly constrain the planet's mass, radius, and orbital parameters.
It is only the third \tess\ giant planet with $e > 0.3$~\citep{Jordan2020,Rodriguez2019}, and it has one of the most eccentric orbits reported to date for a warm Jupiter.

We show that its dynamical state is not consistent with a high-eccentricity migration scenario that would eventually result in the planet becoming a hot Jupiter.
Instead, a past interaction with an undetected massive body has likely caused the planet's extreme orbit.
This valuable addition to the small sample of known warm Jupiter-hosting systems can help constrain the enigma of their origin.
Through its eccentric orbit and the subsequent varying radiative forcing, the planet further holds the promise of observing potential disequilibrium processes in its atmosphere.

This paper is organized as follows:
In Section~\ref{sec:obs} we present the observational data used in this study.
Section~\ref{sec:ana} covers the analysis of these data and concludes with determining properties of the planetary system and its host star.
In Section~\ref{sec:discussion} we discuss implications of our findings and put \plname \, in context with the known exoplanet population.
Finally, in Section~\ref{sec:conclusion} we summarize the results of our study.

We make the code used in the analysis that led to our results available in a public git repository\footnote{\url{https://github.com/matiscke/eccentricWarmJupiter}}.

\section{Observations}\label{sec:obs}
\subsection{\tess\ photometry}\label{subsec:tessphot}
For identifying warm Jupiter candidates, we generated light curves for all bright stars of the TICv8 catalog from the Full Frame Images \citep[FFIs,][]{Jenkins2016} of \tess\ using the \texttt{tesseract}\footnote{\url{https://github.com/astrofelipe/tesseract}} package (Rojas, in prep.).
Briefly, \texttt{tesseract} receives any TIC ID or coordinate as input and performs simple aperture photometry on the FFIs via the \texttt{TESSCut}~\citep{TESSCut} and \texttt{lightkurve} \citep{lightkurve} packages.
Aperture selection was done following~\cite{K2P2}.
Specifically, 293253 and 479184 light curves of bright objects ($\mathrm{T} < 14\,\mathrm{mag}$) have been generated from Sectors 1 and 2, respectively.
For identifying warm Jupiter candidates, we apply a simple algorithm that goes through the light curve searching for zones that significantly deviate in the negative direction from the median flux around a given region.
Then we check by visual inspection if these zones are consistent with a transit-like feature. This procedure allows us in principle to identify also single transiters in a given TESS Sector \citep[e.g.,][]{gill:2020}.
By using this algorithm we found that the star \stname\ presented transit-like periodic features in the two first Sectors of TESS.
An initial fit to the photometric data indicated a period of P$\approx$15.17 and a transit depth of $\delta_\mathrm{TESS}\approx0.8\%$, consistent with the properties of a warm giant candidate given the parameters of the star according to the TICv8 catalog.
\rev{The \tess\ light curve of \stname\ is shown in the upper panel of Fig.~\ref{fig:photometry}.}

\subsection{Photometric follow-up with CHAT}
Due to the limited angular resolution of \tess, ground-based photometry is required to reject false positive scenarios like blended eclipsing binaries.
\stname\ was observed on the night of December 12, 2019 with the 0.7m Chilean-Hungarian Automated Telescope\footnote{\url{https://www.exoplanetscience2.org/sites/default/files/submission-attachments/poster_aj.pdf}} (CHAT) installed at the Las Campanas Observatory.
Observations were performed with a Sloan $i^{\prime}$ photometric filter using a mild defocus and exposure times of 110 sec. 
We processed the data with a dedicated pipeline developed to produce high precision light curves using differential photometry \citep[e.g.][]{espinoza:2019,jones:2019} with the LCOGT 1.0m telescopes~\citep{Brown2013}.
The optimal photometric precision was obtained with an aperture of 14~pixels (8.3$\arcsec$).
We plot the obtained light curve in the bottom right panel of Fig.~\ref{fig:photometry}.
We recovered a full transit, which confirms that the transit-like features identified in the \tess\ data occur in \stname.
\rev{The transit depth of $\delta_\mathrm{CHAT} = 0.0087\pm0.0004$ is consistent with the signal identified in the \tess\ photometry.}

\subsection{Additional photometry from LCOGT}
Because of a rather grazing transit configuration, the posterior probability densities from our initial fits contained a strong degeneracy between the scaled planetary radius $R_\mathrm{P}/R_\star$ and the impact parameter $b$.
To lift this degeneracy and to improve the constraint on the planet radius, we obtained additional transit photometry with the Las Cumbres Observatory Global Telescope (\textit{LCOGT}) Network on July~16,~2020 (see Fig.~\ref{fig:photometry}).
The measurements were taken in the $i^{\prime}$ band and cover all phases of the transit.
To maximize photometric precision, we chose an aperture of 24~pixels (9.4$\arcsec$).
\rev{We recover a transit depth of $\delta_\mathrm{LCOGT} = 0.0083\pm0.0002$. Within the uncertainties, this is consistent with the values obtained for the other instruments.}
Including the additional data in the fit leaves only little residual correlation between $R_\mathrm{P}/R_\star$ and $b$ and strongly improved the posterior on $R_\mathrm{P}$~(see Sect.~\ref{sec:jointFit}).

We make all our follow-up light curves available on \mbox{exoFOP}\footnote{\url{https://exofop.ipac.caltech.edu/tess}}.

\subsection{High precision spectroscopy with FEROS}
We obtained high-resolution ($R=48000$) spectra with the Fiber-fed Extended Range Optical Spectrograph \citep[FEROS, ][]{kaufer:99}, mounted at the \SI{2.2}{\meter}~MPG telescope at La Silla Observatory.
In total, 25 exposures of \SI{1200}{\second} were taken between June~19,~2019 and March~9,~2020.
From these, we extracted radial velocities (RV) using the \texttt{CERES} pipeline \citep{brahm:2017:ceres}, which performs all steps from bias, dark, and flat-field calibration to cross-correlation matching of the resulting spectrum with a G2-type binary mask.
The observations were performed in the simultaneous calibration mode for tracking the instrumental velocity drift produced by changes in the spectrograph environment. This procedure involves the monitoring of a ThAr spectrum with a second fiber. The typical signal-to-noise ratio of these spectra was about 70.
The time series of FEROS RV measurements is shown in Fig.~\ref{fig:RV} and they are listed in Table \ref{tab:RVdata}.

\subsection{Contamination}
We checked for possible closeby sources that could contaminate our photometric aperture with their light.
Any sources within $\sim \SI{10}{\arcsecond}$, which is the photometric aperture we used for our \textit{LCOGT} photometry, could cause such contamination.
After querying the GAIA DR2 catalog~\citep{gaia2018} we found the closest source to \stname\ at an angular separation of $\sim \SI{46}{\arcsecond}$.
We thus find no evidence for significant contamination of our photometry.

\section{Analysis} \label{sec:ana}

\subsection{Stellar Parameters}\label{sec:stellar}
For characterizing the host star, we first determined its atmospheric parameters from the co-added FEROS spectra. Specifically, we used the \zaspe\ code \citep{brahm:2016:zaspe} which compares the observed spectrum against a grid of synthetic ones generated from the ATLAS9 model atmospheres \citep{atlas9}. We then used the PARSEC evolutionary models \citep{bressan:2012}, as described in \citet{hd1397}, to determine the physical parameters of the star.
Briefly, we compared the observed broad band photometric magnitudes of the star with those generated with models having different physical parameters by taking into account the distance determined from the Gaia~DR2 parallax and assuming an extinction law of \citet{cardelli:89} dependent on the A$_V$ parameter.
The parameter space was explored using the \texttt{emcee} \citep{emcee:2013} package.
The obtained atmospheric and physical parameters of \stname\ are listed in Table \ref{tab:stprops}. \stname\ is a main sequence G-type star with a mass of \mstar\ = \mst\ \msun, a radius of \rstar\ = \rst\ \rsun, and an age of \age\ Gyr.
\stname\ is slightly metal rich ([Fe/H] = \fehv\ dex) and has an effective temperature of \teff\ = \teffv\ K.
We note that the quoted uncertainties do not account for possible systematic errors in the stellar evolutionary models.

\begin{deluxetable}{lrc}[b!]
\tablecaption{Stellar properties of \stname\  \label{tab:stprops}}
\tablecolumns{3}
\tablewidth{0pt}
\tablehead{
\colhead{Parameter} &
\colhead{Value} &
\colhead{Reference} \\
}
\startdata
Names \dotfill   &    \stname  &  \tess \\
 & 2MASS J01294694-6044238 & 2MASS  \\
 & UCAC4 147-001388 & UCAC 4  \\
RA \dotfill (J2015.5) &  01h29m46.99s &  GAIA\\
DEC \dotfill (J2015.5) & -60d44m23.67s &   GAIA\\
pm$^{\rm RA}$ \hfill (mas yr$^{-1}$) & 18.053 $\pm$ 0.036& GAIA\\
pm$^{\rm DEC}$ \dotfill (mas yr$^{-1}$) & 10.523 $\pm$ 0.034 & GAIA\\
$\pi$ \dotfill (mas)& 3.23 $\pm$ 0.02 & GAIA \\
\hline
T \dotfill (mag) & 11.486 $\pm$ 0.006 & \tess\\
B  \dotfill (mag) & 12.746 $\pm$ 0.015 & APASS\\
V  \dotfill (mag) & 12.144 $\pm$ 0.069 & APASS\\
J  \dotfill (mag) & 10.858 $\pm$ 0.023 & 2MASS\\
H  \dotfill (mag) & 10.571 $\pm$ 0.024 & 2MASS\\
K$_s$  \dotfill (mag) & 10.485 $\pm$ 0.021 & 2MASS\\
WISE1  \dotfill (mag) & 10.463 $\pm$ 0.023 & WISE\\
WISE2  \dotfill (mag) & 10.518 $\pm$ 0.021 & WISE\\
WISE3  \dotfill (mag) & 10.408 $\pm$ 0.059 & WISE\\
\hline
\teff  \dotfill (K) & \teffv & this work\\
\logg \dotfill (dex) & \loggv & this work\\
\feh \dotfill (dex) & \fehv & this work\\
\vsini \dotfill (km s$^{-1}$) & \vsiniv & this work\\
\mstar \dotfill (\msun) & \mst & this work\\
\rstar \dotfill (\rsun) & \rst & this work\\
L$_{\star}$ \dotfill (L$_{\odot}$) & 1.196 $\pm$ 0.050 & this work\\
Age \dotfill (Gyr) & \age & this work\\
A$_V$ \dotfill (mag) & $0.117_{-0.063}^{+0.068}$ & this work\\
$\rho_\star$ \dotfill (g cm$^{-3}$) & \rhost  & this work\\
\enddata
\end{deluxetable}

\subsection{Joint modeling}
For the joint photometry and RV modeling, we employed the python package \texttt{juliet}~\citep{juliet}.
This tool uses existing codes to model transit photometry~\citep[\texttt{batman},][]{kreidberg:2015} and radial velocity time series \citep[\texttt{radvel},][]{fulton:2018}.
It further allows us to incorporate Gaussian Process Regression (GP) via the \texttt{celerite} package~\citep{Foreman-Mackey2017}, which we employ to model systematic nuisance signals.
To explore the parameter space, it uses the \mbox{MultiNest} nested sampling technique \citep{MultiNest}, implemented in the \texttt{pyMultiNest} software package \citep{Buchner2014}.
\texttt{juliet} further calculates evidences $Z_i = \mathrm{P}(M_i(\theta)|\mathcal{D})$ for models $M_i$ with sets of parameters $\theta$ given the data $\mathcal{D}$.
To compare two models $M_i, M_j$, we compute the differences of their log-evidences,
\begin{equation}
    \Delta \ln Z_{i, j}= \ln Z_i/Z_j = \ln \left[\mathrm{P}(M_i(\theta) | \mathcal{D})/ \mathrm{P}(M_j(\theta) | \mathcal{D}) \right].
\end{equation}
Here, we adopted a general rule of thumb that if $\Delta \ln Z_{i, j}\geq 3$, the model with the larger log-evidence is favored.
If $\Delta \ln Z_{i, j} \lesssim 3$, we consider the models to be indistinguishable and prefer the simpler one.
As the MultiNest algorithm is known for showing scatter in $\ln Z$ that exceeds the reported uncertainties~\citep[e.g.,][]{Nelson2020}, we always repeated the calculations several times.
The variations among such runs were always smaller than one and therefore negligible for our purposes.

\subsubsection{Model parameters}
For several inferred quantities, we fitted parametrizations that allow for efficient sampling and are limited to physically plausible values:
\begin{itemize}

  \item \textbf{Limb darkening coefficients}: to ensure uniform sampling of only physical solutions, we used a triangular sampling scheme. As outlined in \citet{Kipping2013a}, we transformed the quadratic limb darkening coefficients $u_1, u_2$ to $q_1 = (u_1 + u_2)^2$ and $q_2 = 0.5 u_1(u_1 + u_2)^{-1}$. For ground-based photometry, we assumed a linear limb darkening profile and $q_1 = u_1$.
    \item \textbf{Prior for the stellar density $\rho_\star$}: from our stellar modeling (Sect. \ref{sec:stellar}), we obtained a distribution for the stellar density $\rho_\star$ which we used as a prior for our joint fit instead of the scaled semi-major axis of the planetary orbit.
    \item \textbf{Eccentricity and argument of periastron}: we parameterized the orbital eccentricity $e$ and the argument of periastron $\omega$ as  $\mathcal{S}_1 = \sqrt{e} \sin \omega$ and $\mathcal{S}_2 = \sqrt{e} \cos \omega$ and ensure at each iteration that $e = \mathcal{S}_1^2 + \mathcal{S}_2^2 \leq 1$.
\end{itemize}

 Given the observed RV variations and empirical mass-radius relationships \citep[e.g.,][]{Chen2016, Neil2019}, it is justified to neglect extreme radius ratios.
 	We thus constrained the sampling to $\rpl/\rstar < 0.5$.

\subsubsection{Limb darkening}
The limb darkening profile of \stname \, is poorly constrained by our available data; we therefore simultaneously fit for it in the joint fit.
An optimal choice of a limb darkening law is not straight-forward: there is a trade-off between accuracy and computational cost, and the performances of different laws depend on the noise level of the light curve (see \citet{Schlecker2016} for a more detailed discussion).
To account for the different noise levels in space-based and ground-based photometry \citep{Espinoza2016f}, we decided to use a quadratic limb darkening law for \tess\ photometry and a linear law for the \textit{CHAT} and \textit{LCOGT} light curves.

\subsubsection{RV analysis}	

\begin{figure*}
\gridline{\fig{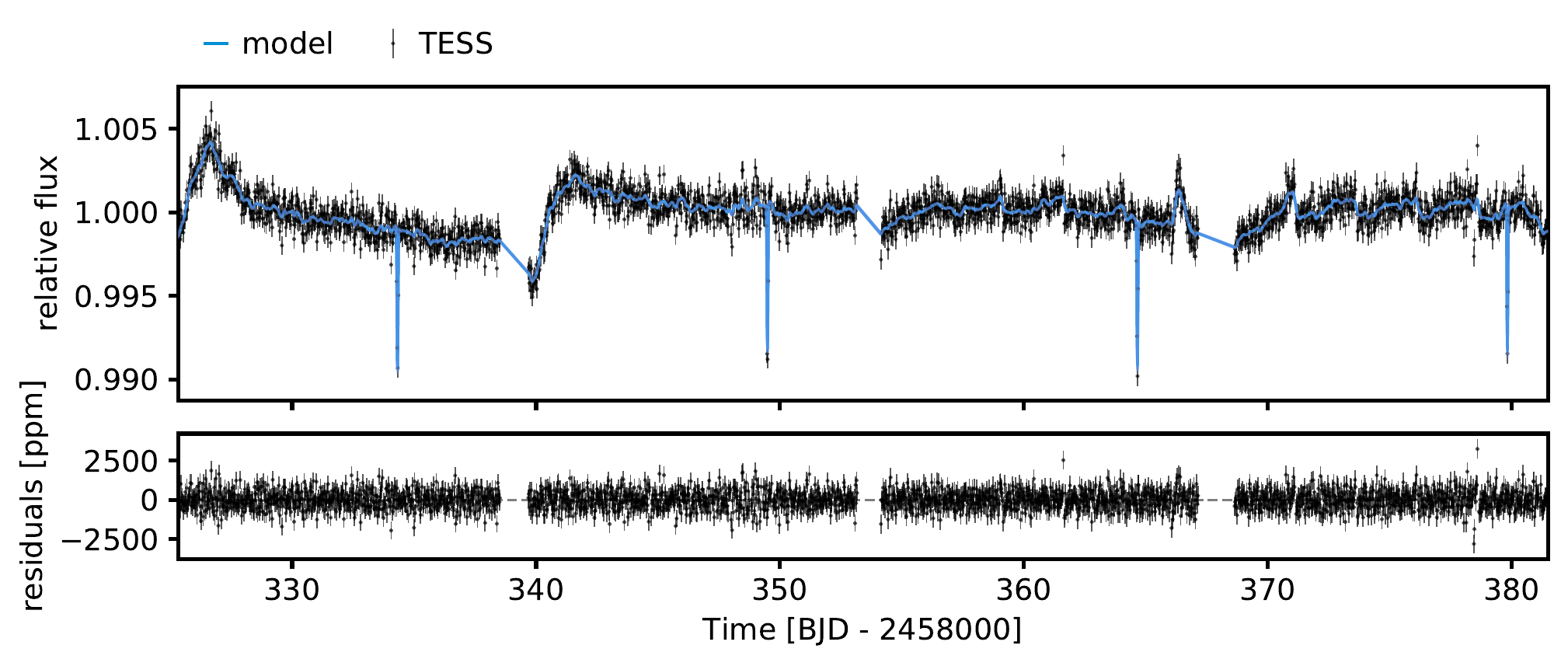}{0.98\textwidth}{}}
\gridline{\fig{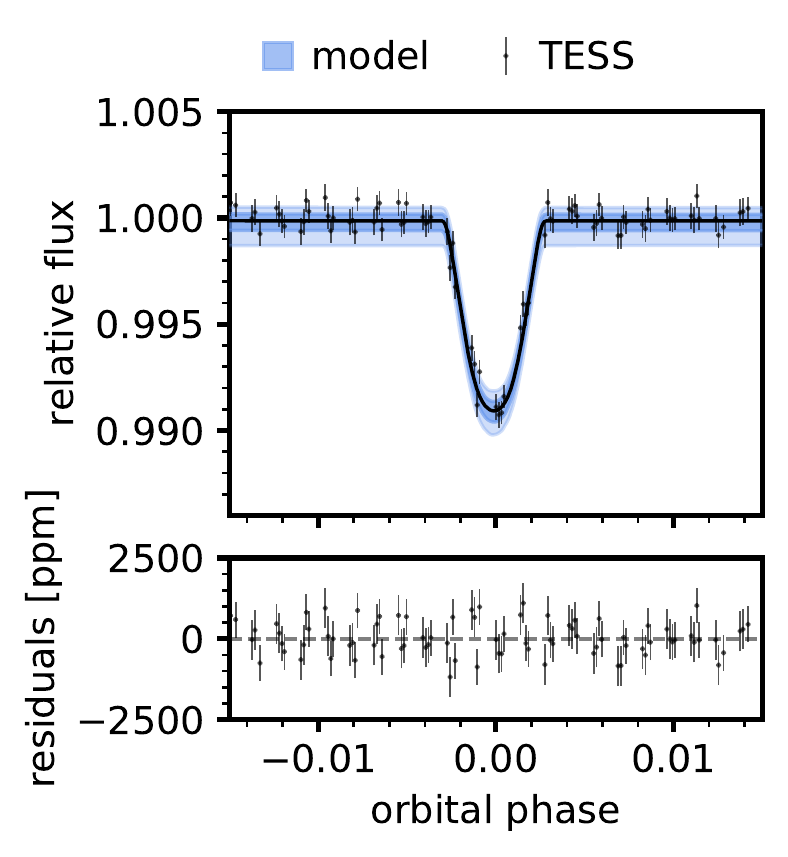}{0.345\textwidth}{}\fig{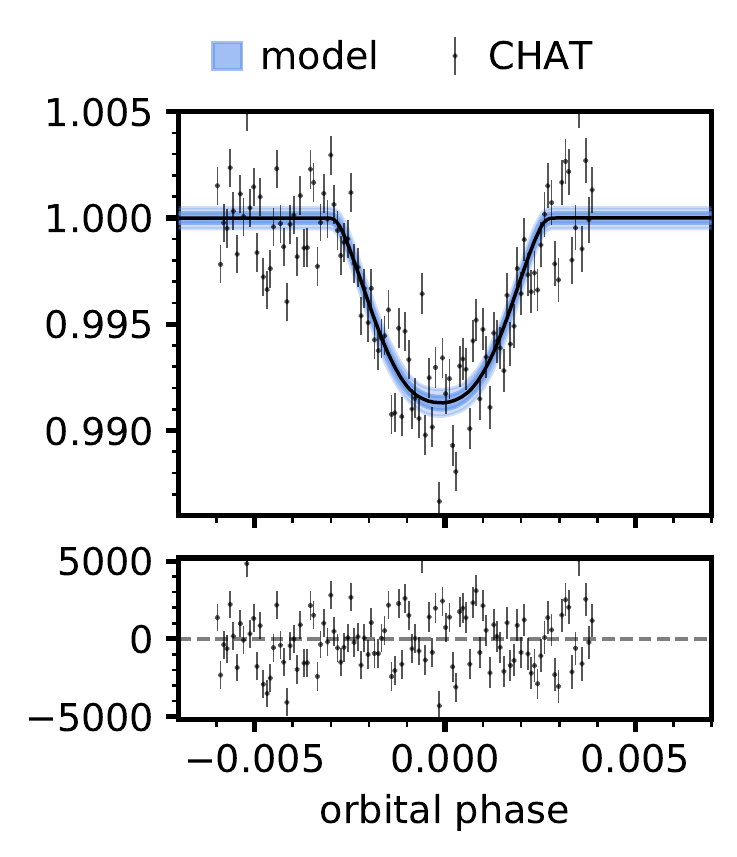}{0.32\textwidth}{}\fig{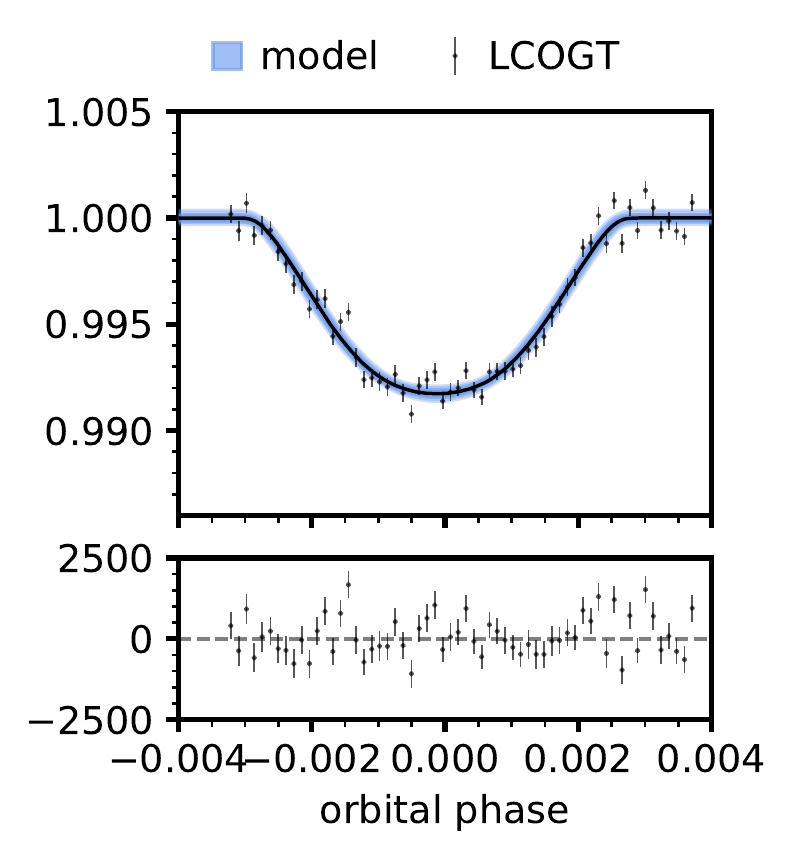}{0.34\textwidth}{}}
\caption{Photometry for \stname.
Gray points represent the relative flux and errors. Solid lines show the theoretical light curve using the best-fit parameters derived in the joint modeling including GP. Blue shaded regions denote the \SI{68}{\percent} and \SI{95}{\percent} credibility bands of the model.
Residuals are shown below each light curve.
\textit{Top:} Full \tess\ light curve generated from 30-minute-cadence photometry of Sectors 1 and 2.
\textit{Bottom left:} Phase-folded \tess\ photometry around the transit events.
\textit{Bottom center:} Follow-up photometry of a single transit obtained with \textit{CHAT} in the $i^{\prime}$ band.
\textit{Bottom right:} \textit{LCOGT} photometry of a single transit ($i^{\prime}$ band). This additional transit photometry lifted the $R_\mathrm{P}/R_\star - b$ degeneracy and strongly improved our constraint on the planet radius.}
\label{fig:photometry}
\end{figure*}

\begin{figure*}
\gridline{\fig{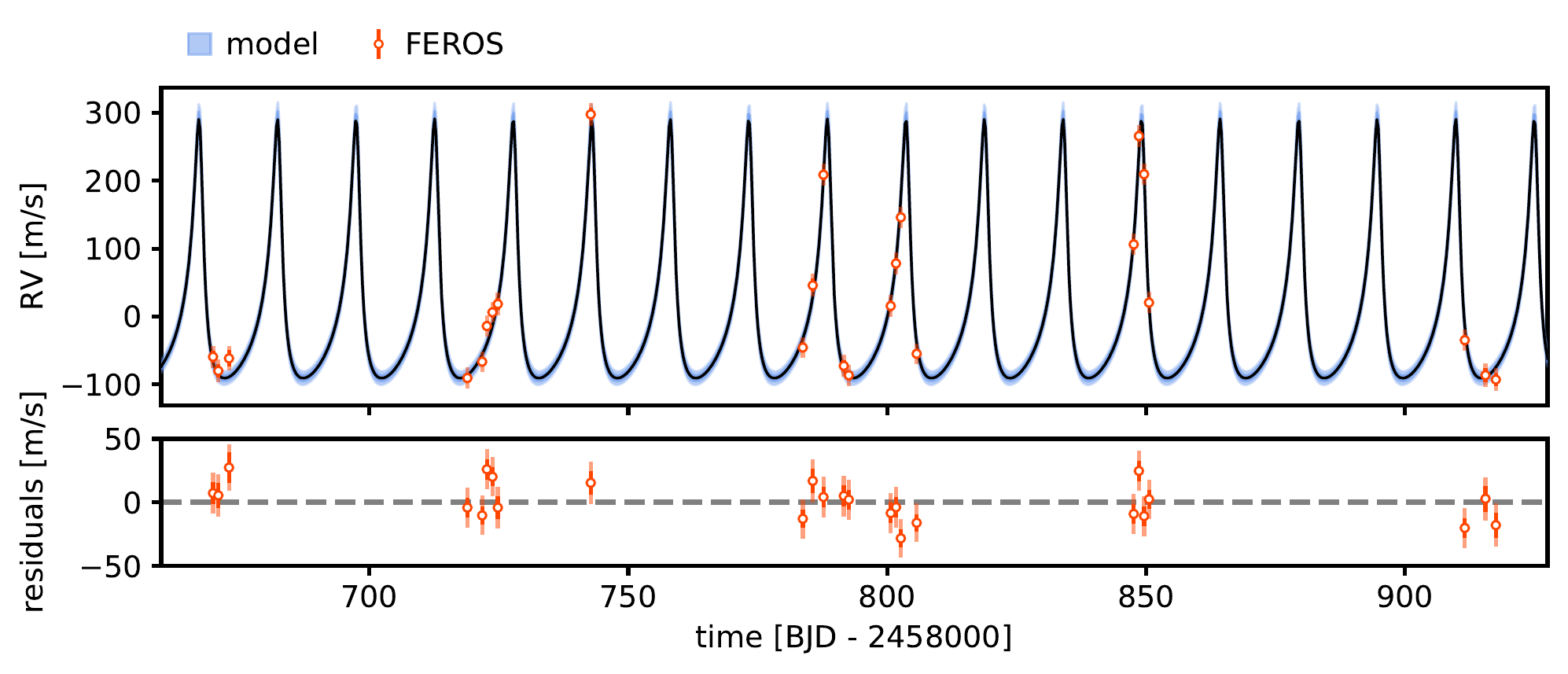}{0.99\textwidth}{}}
\gridline{\fig{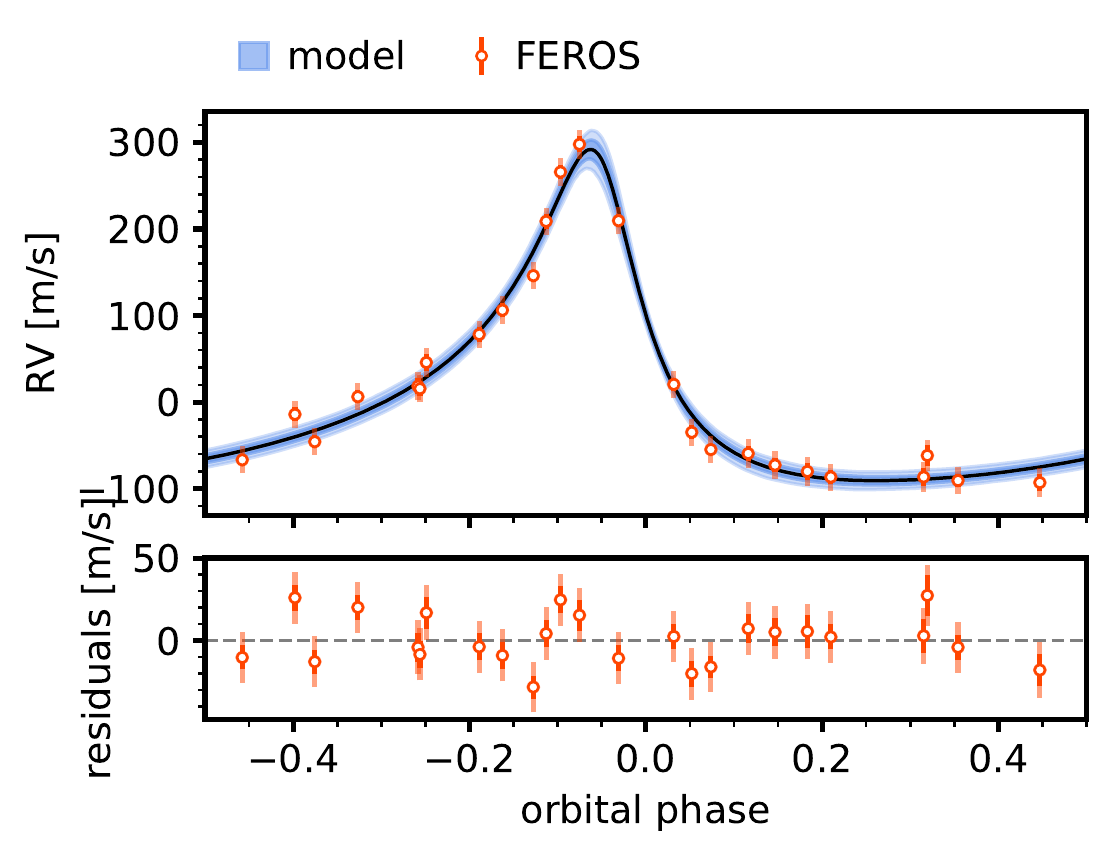}{0.48\textwidth}{}\fig{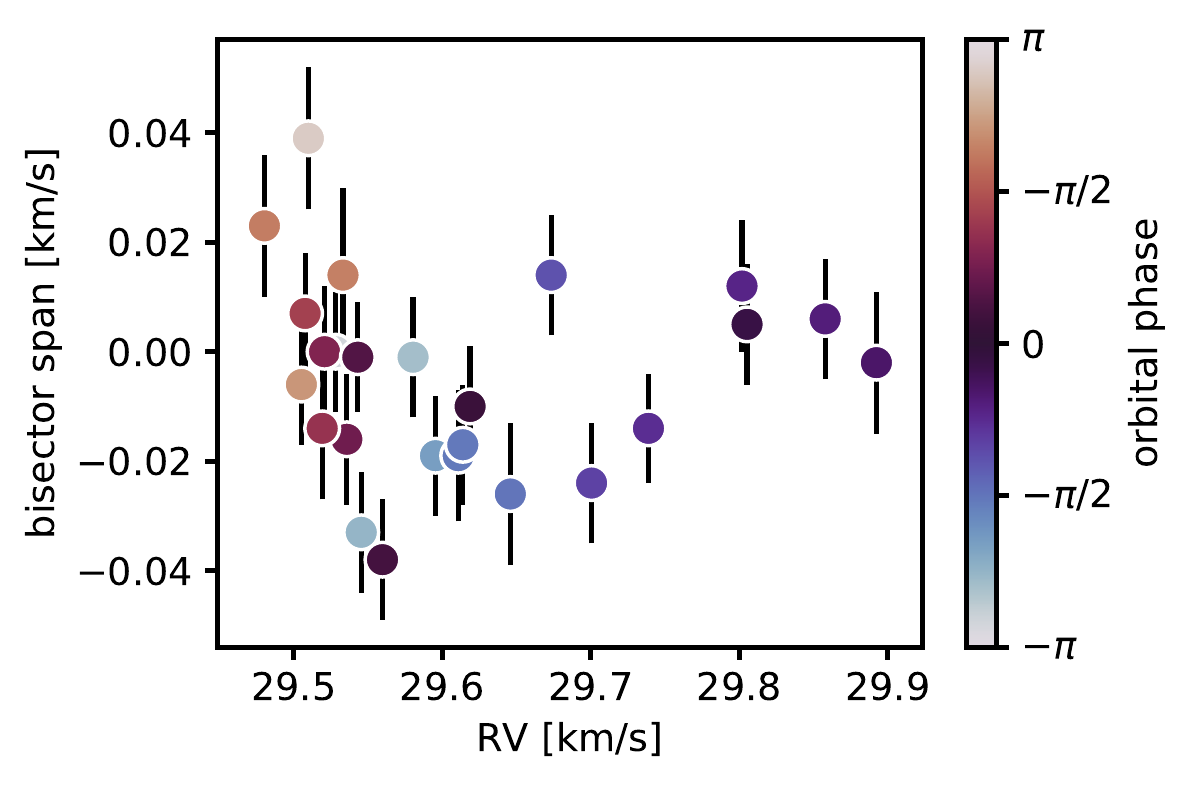}{0.52\textwidth}{}}
\caption{Radial velocity time series for \stname.
Light error bars reveal the best-fit jitter term, which we added in quadrature to the RV errors.
The model generated with the derived parameters of our joint modeling is plotted with a black line, and blue bands denote its \SI{68}{\percent} and \SI{95}{\percent} posterior credibility intervals.
Residuals are obtained by subtracting the median posterior model from the data.
\textit{Top:} RV time series measured with the FEROS spectrograph.
\textit{Left}: phase-folded RV measurements obtained with FEROS.
\textit{Right}: bisector span as a function of radial velocity. The color of each measurement represents the orbital phase at which it was taken, assuming our best-fit period.}
\label{fig:RV}
\end{figure*}

\begin{figure}
\epsscale{1.2}
\plotone{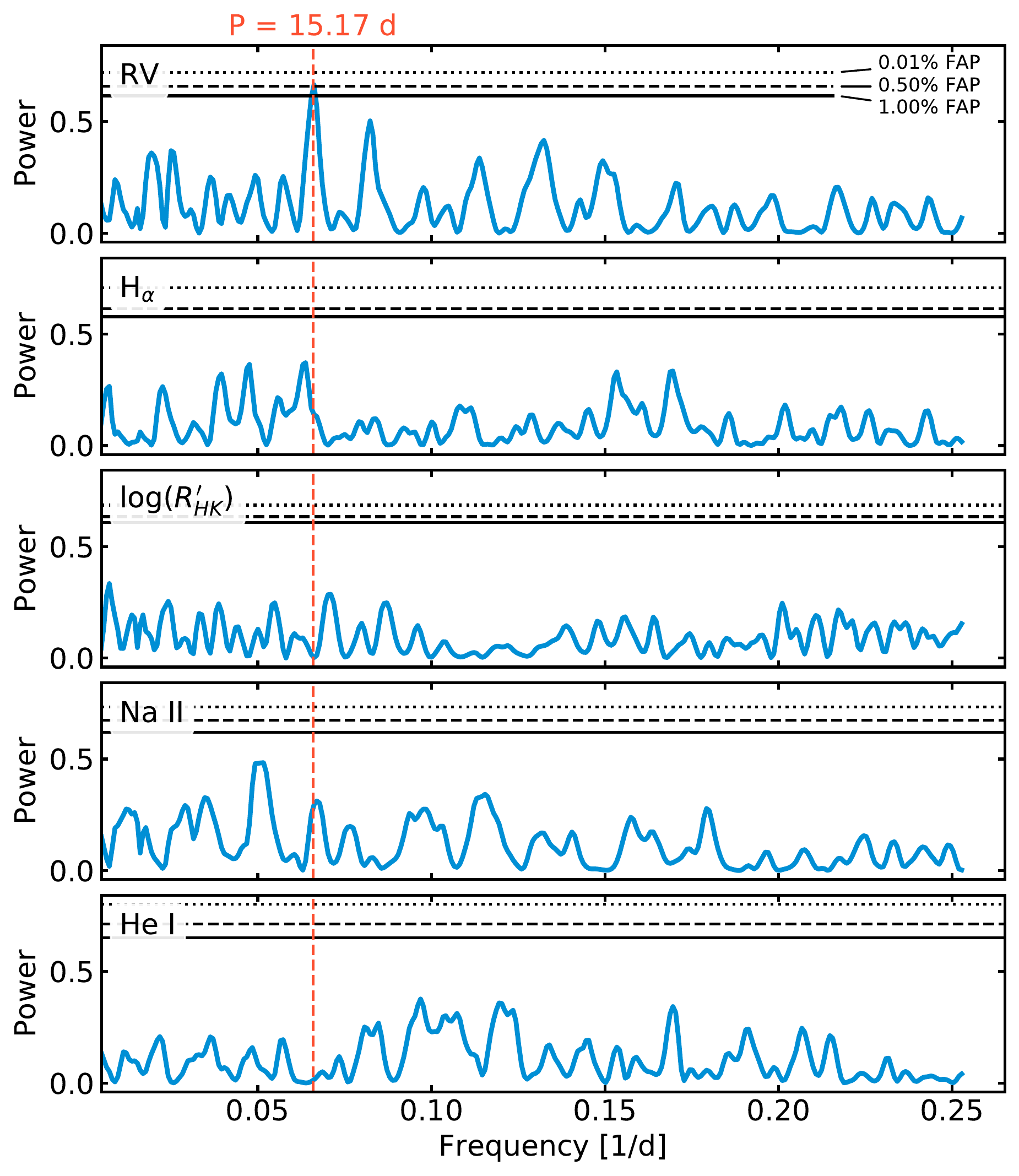}
\caption{Generalized Lomb-Scargle periodograms of radial velocity time series and common activity indicators. Solid, dashed, and dotted lines mark 1\,\%, 0.5\,\%, and 0.01\,\% false alarm probabilities, which we computed via bootstrap resamplings.
The orange line marks the orbital period of \plname.}
\label{fig:periodograms}
\end{figure}

The RV time series show a strong signal with a period corresponding to the candidate transiting planet ($P=\SI{15.17}{\day}$, see Fig.~\ref{fig:RV}, Fig.~\ref{fig:periodograms}).
To assess the evidence of this signal being of planetary origin, we compared models with and without a planet based on only the FEROS RV dataset.
We further evaluated models including more than one planet and compared the log-evidences of the different cases:
\begin{enumerate}
    \item \textbf{No planet}: we assumed that all RV variations are due to astronomical and instrumental ``jitter''. The only free parameters were $\mu_\mathrm{FEROS}$ and $\sigma_\mathrm{FEROS}$ using the same prior distributions as in Table~\ref{tab:priors}.
    This ``flat'' model resulted in a log-evidence $\ln(Z) = -161$.
    \item \label{enum:1pl} \textbf{Single planet}: we assumed there is a planetary signal in the RV data and widened the orbital period prior to a uniform distribution of \SIrange{1}{30}{\day}.
    The RV semi-amplitude $K$ was free to vary between zero and \SI{1000}{\meter\per\second} (uniform prior).
    For $T_0$, we chose a uniform prior ranging from the first photometric observation to \SI{124}{\day} later, which corresponds to half the RV baseline.
    All other free parameters had the same priors as in our nominal model.
    This fit converged to a similar solution as our final model with a period distribution consistent with the intervals between the observed transit events. $\ln(Z)=-135$.
    \item \label{enum:1plcirc} \textbf{Single planet, circular orbit}: same as~\ref{enum:1pl}., but fixing the eccentricity to zero (and $\omega$ to an arbitrary \SI{90}{\degree}).
    The fit converged to a solution with $P$ similar to the period distribution in our final joint fit, but the jitter term is strongly increased to account for the large RV variations.
    With $\ln(Z)=-160$, the evidence of this model is similar to the one belonging to the no-planet hypothesis.
    \item \label{enum:2plcirc} \textbf{Two planets, circular orbits}: same as~\ref{enum:1plcirc}., but assuming a second planet in the system.
    For this hypothetical additional planet, we let the orbital period vary within \SIrange{1}{30}{\day} and used the same uniform prior $\mathcal{U}(0, 1000)\,\SI{}{\meter\per\second}$ for its RV semi-amplitude $K_2$.
    The $\sim \SI{15}{\day}$ periodicity is recovered, but no stable solution in favor of a two-planet-scenario is evident.
    $\ln(Z)=-164$
    \item \textbf{Two planets, circular and eccentric orbits}: same as~\ref{enum:2plcirc}., but one planet with freely varying eccentricity.
    The eccentric, \SI{15}{\day} candidate signal is recovered.
    The period and RV semi-amplitude of the second planet are poorly constrained.
    $\ln(Z)=-151$
    \item \label{enum:2plecc} \textbf{Two planets, both on eccentric orbits}: same as~\ref{enum:2plcirc}., but with free eccentricity for both planets.
    Again, the \SI{15}{\day} signal is strongly recovered, while the weak signal of an additional planet is poorly constrained.
    $\ln(Z)=-144$
\end{enumerate}
We list all model evidences in Table~\ref{tab:lnZ_RV}.
The log-evidence difference between the preferred model (\ref{enum:1pl}. Single planet) and the runner-up (\ref{enum:2plecc}. Two planets, both on eccentric orbits) $\Delta \ln Z_{6, 2} \approx 9$, which corresponds to a Bayes factor of $\sim 10^4$.
The difference to the flat model is as large as $\Delta \ln Z_{1, 2} \approx 26$, implying a Bayes factor of $\sim 10^{11}$.
Thus, the planetary model is strongly favored above the flat model and an eccentric single-planet solution is preferred.

To test if the candidate signal could potentially be associated with stellar activity, we produced generalized Lomb-Scargle periodograms~\citep{Zechmeister2009a} for the radial velocity time series, as well as for common activity indicators based on FEROS data (Fig.~\ref{fig:periodograms}). In particular, we obtained the H$_\alpha$, log($R^\prime_{HK}$), Na~II, and He~I activity indices, which trace chromospheric activity. We computed H$_\alpha$ following \citet{Boisse09}. As \stname\ is a G-type star, we used the regions defined by \citet{Duncan91} and the calibrations of \citet{Noyes84} for log($R^\prime_{HK}$). For Na~II and He~I we followed \citet{Gomes11}.
For each time series (RV, H$_\alpha$, log($R^\prime_{HK}$), Na~II, and He~I), we computed the power levels for 1\,\%, 0.5\,\%, and 0.01\,\% false alarm probabilities (FAP) by a bootstrap method and plot them as solid, dashed, and dotted lines, respectively. There is a strong signal in the periodogram of the radial velocities at the 15.17d period, below 0.5\,\% FAP. Meanwhile, there are no significant signals visible in the periodograms of any of the activity indices, indicating that the radial velocity signal is unlikely to come from quasi-periodic stellar activity.

To further show beyond doubt that the measured radial velocity variations represent reflex motions of the star, we also tested if they can be caused by variations in the stellar photosphere.
A well-established method to do this is the inspection of atmospheric line profiles, which should be constant in time for actual stellar velocity changes.
Specifically, the bisector span (BIS) can serve as a diagnostic to search for possible false positive scenarios~\citep[e.g.,][]{Queloz2001a}.
We are interested in its correlation with the RV time series and orbital phase, and confront these variables in the right panel of Fig.~\ref{fig:RV}.
Here, we plot the FEROS RV measurements against their bisector spans.
The points are color-coded by the orbital phase of the measurements, where zero phase is at $T_0 + n P$ using median values from our nominal fit.
While bisector span and RV show no evidence for correlation, this may not be true for bisector span and orbital phase.
A Spearman's rank coefficient of $0.45^{+0.31}_{-0.39}$, where we quote \SI{95}{\percent} confidence intervals from a bootstrapped sampling method\footnote{\texttt{pingouin.compute\_bootci} from the \texttt{Pingouin} python package~\citep{Vallat2018}}, permits the suspicion of a positive correlation.
However, due to the small number of data points we cannot reject the null hypothesis that there is no monotonic association between the two variables.
In addition, the bisector span variations are on the order of \SI{10}{\meter\per\second} and cannot account for the observed RV \rev{semi-amplitude of $K\approx\SI{191}{\meter\per\second}$.}
The line profiles thus provide further evidence that the observed RV variations are indeed due to velocity changes of the target star and not caused by atmospheric variations.

In addition to the tests described above, the non-sinusoidal pattern of the RVs is a strong indication for orbital motion as opposed to stellar activity.
We conclude that the radial velocity time series independently confirms the planet hypothesis.
In the following, we refer to the confirmed exoplanet as \plname.\footnote{\rev{We submitted the object to exoFOP as a community TESS Object of Interest (CTOI); it is now listed as TOI~2179.01.}}

\subsubsection{Joint photometry and RV fit}\label{sec:jointFit}
We performed a simultaneous fit on photometric and spectroscopic datasets (\textit{TESS, CHAT, LCOGT}, and \textit{FEROS}) to jointly constrain all planetary parameters of \plname.
\rev{To account for the long cadence in the \textit{TESS} light curve, we modeled the transits with 20-fold supersampling using the exposure time of the actual observations.}
The initial photometric fit (see Sect.~\ref{subsec:tessphot}) provided narrow constraints on the orbital period $P$ and time of mid-transit $T_0$.
We used the median values obtained there to construct Gaussian priors for these parameters, but we enlarged the dispersions (compare Table \ref{tab:priors}).
For the instrument-specific flux offsets $M_{\textnormal{TESS}}$, $M_{\textnormal{CHAT}}$, and $M_{\textnormal{LCOGT}}$ we assumed Gaussian priors based on our photometric fit.
A Gaussian prior on the stellar density was motivated by the analysis of the stellar parameters presented in Sect.~\ref{sec:stellar}.
\rev{Additional confidence for this prior stems from a separate joint fit where we used an uninformative prior ($\mathcal{J}(10^{2},(10^{4})^2)$) on $\rho_\star$.
The resulting posterior probability, $\rho_\star = 996^{+257}_{-421}$, and the result from our stellar analysis agree within the uncertainties.
}
For all other parameters, we chose uninformative priors to sample the whole physically plausible parameter space.

There are potentially time-correlated processes such as instrumental red noise, stellar variability, or blended sources that are not covered by our astrophysical model.
We account for this red noise by adding a Gaussian Process (GP) component to the \tess\ photometry  with an exponential kernel as implemented in the \texttt{celerite} software package \citep{Foreman-Mackey2017}.
This adds two additional hyperparameters: an amplitude $\sigma^{GP}_{\textnormal{TESS}}$ and a timescale $\tau^{GP}_{\textnormal{TESS}}$.
For comparison, we performed the same fit with and without a GP component.
The variant including GP performed significantly better than the white-noise model and we thus consider it our nominal model.

In the same manner, we tested adding a GP component to the \textit{LCOGT}~photometry, which shows possible systematic effects in the residuals~(compare Fig.~\ref{fig:photometry}).
Here, we chose a Mat\'{e}rn~3/2 kernel, which again adds two hyperparameters for timescale and amplitude to the model.
The model including the GP component consistently performed comparable ($|\Delta \ln Z| < 1$) or worse than the model without, which is why we chose to continue with the less complex noise model without GP.

\begin{figure*}
    \epsscale{1.05}
	\centering
    \plotone{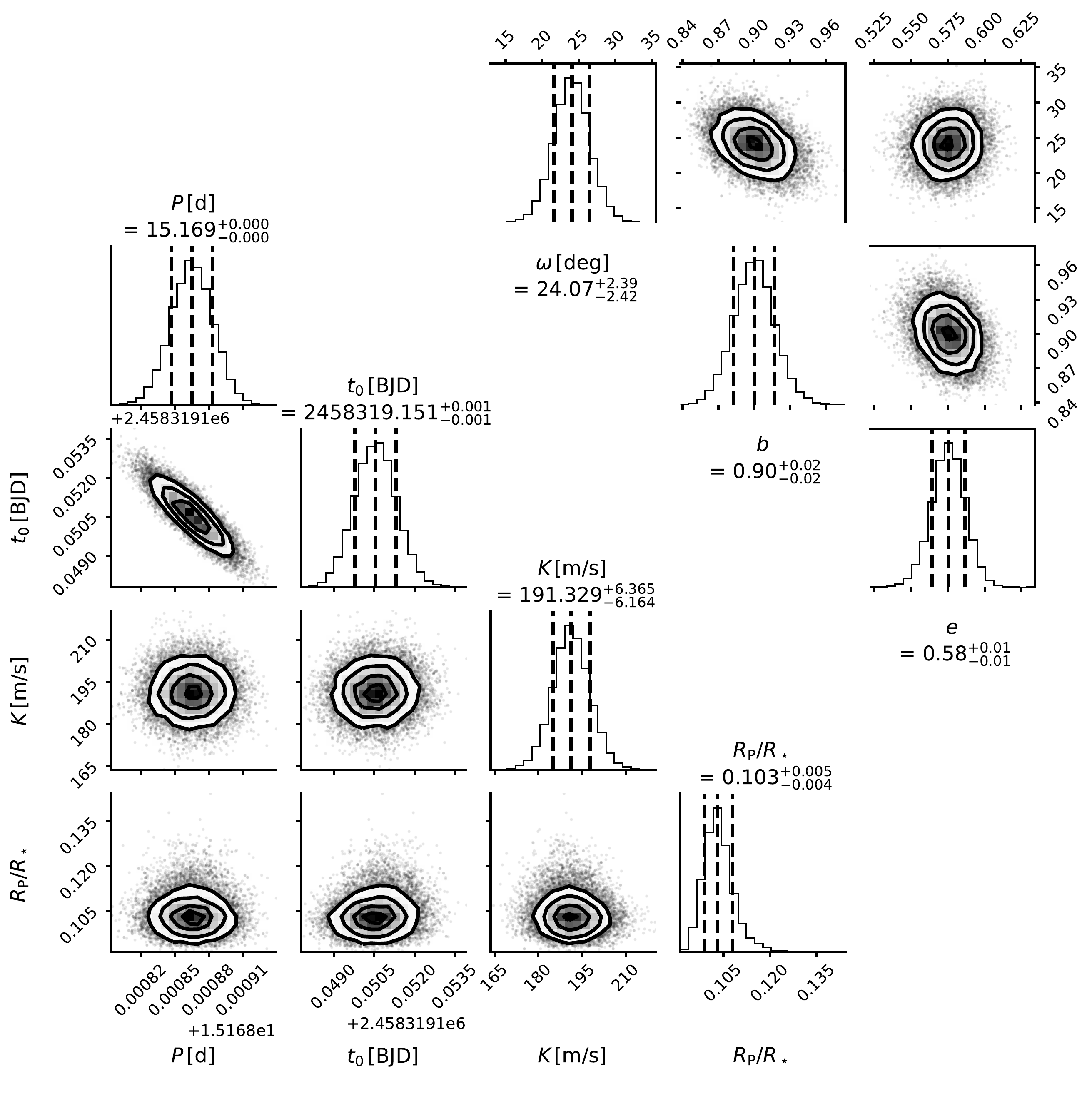}
	\caption{Posterior distributions of planetary parameters. The lower left triangle shows fitting parameters and the upper right triangle shows derived parameters of \plname's orbit. The stated values represent 16th, 50th, and 84th percentiles. See Appendix~\ref{sec:fullCorner} for the full sample and a discussion about correlated parameters.}\label{fig:cornerMajorParams}
\end{figure*}

Figures \ref{fig:photometry} and \ref{fig:RV} show the photometric and radial velocity time series resulting from this model using its median parameters (solid lines).
Dark (light) blue bands show the \SI{68}{\percent} (\SI{95}{\percent}) credibility bands of the model.
The residuals below the time series show the measured data with the median posterior model subtracted; both RV and photometry residuals appear inconspicuous.
Figure~\ref{fig:cornerMajorParams} shows the posterior distributions for the planet's main parameters as sampled in our nominal fit.
All distributions are approximately Gaussian and barely correlated, except for the planet-to-star ratio $R_\mathrm{P}/R_\star$ (see Appendix~\ref{sec:fullCorner} for a discussion).
We present the posterior distributions of the model parameters alongside the derived physical parameters in Table \ref{tab:posteriors}, where we state for each parameter distribution the 16th, 50th, and 84th percentile.
Notably, with a planetary \rev{bulk density $\rho_\mathrm{P}~\approx~1847~\,~\mathrm{kg \, m^{-3}}$,} \plname's average density is comparable to Neptune's.
By sheer coincidence, the planet's period and eccentricity resemble that of the \tess\ spacecraft (\SI{13.7}{\day}, $e = 0.55$)~\citep{Ricker2014}.

\subsubsection{Search for additional planets}
We repeated the joint fit with an additional linear RV term to search for any long-period companions that would locally cause a linear trend in the RVs.
To this end, we include intercept and slope parameters with wide, normal priors for another joint fit.
The result is consistent with an RV slope of zero and the log-likelihood of the model including the linear trend is suppressed with $\Delta \ln Z \approx 7$.
We conclude that the data at hand does not support additional outer companions in the system.

There is also no evidence of interior planets, which is expected since the deep intrusion of \plname\ into the inner system leaves only limited room for stable inner orbits~\citep[e.g.,][]{Gladman1993}.
In fact, planets like \plname\ have been suggested to be a main cause for the destruction of inner systems of low-mass planets~\citep{Schlecker2020}.

\begin{deluxetable}{lrr}[b!]
\tablecaption{Model evidences from RV fits for different models. $\Delta \ln Z$ states the difference in log-evidence compared to our best model ``1~planet, eccentric orbit''.\label{tab:lnZ_RV}}
\tablecolumns{3}
\tablewidth{\textwidth}
\tablehead{
\colhead{Model} &
\colhead{log-evidence $\ln Z$} &
\colhead{$\Delta \ln Z$}\\
}
\startdata
0 planets & -160.94 $\pm$  0.14 & -26.41\\
1 planet, eccentric orbit & -134.53 $\pm$  0.02 & 0\\
1 planet, circular orbits & -160.46 $\pm$  0.04 & -25.93\\
2 planets, circular orbits & -164.34 $\pm$  0.02 & -29.81\\
2 planets, circular\&eccentric orbits & -150.76 $\pm$  0.01 & -16.23\\
2 planets, eccentric orbits & -144.46 $\pm$  0.01 & -9.93\\
\hline
\enddata
\end{deluxetable}

\subsection{Approximation of the planetary equilibrium temperature}\label{sec:Teq}

The equilibrium temperature $T_\mathrm{eq}$ that the planet maintains if it is in energy balance with the radiation input from the host star is a determining factor for the physical properties of its atmosphere.
Due to \plname's considerable eccentricity, this input is not constant over its orbit.
To give a first-order estimate on the temperature range that the planet can assume, we investigated two extreme cases of planetary heat adjustment:
\begin{enumerate}
	\item instantaneous heat adjustment ($T_\mathrm{eq,inst}$). Here, we assumed that the planetary atmosphere adjusts to the changing irradiation without any time delay.
For this situation, we used the approximation in \citet[][Equation 3]{Kaltenegger2011}.
    \item orbitally averaged heat adjustment ($T_\mathrm{eq,avg}$).
In this case, the planetary temperature remains in equilibrium with the incoming stellar energy, i.e. $T_\mathrm{eq} = \mathrm{const.}$ over one orbit.
    To approximate this temperature, we used a temporal average for elliptic orbits~\citep[][Equation 16]{Mendez2017}.
\end{enumerate}
For both extremes, we assumed that the heat flux from the planet's interior is negligible compared to the stellar irradiation and ignored any internal energy sources.
The infrared emissivity $\varepsilon_\mathrm{IR}$ was fixed to unity.
We further assumed two cases for how atmospheric flow distributes the incoming stellar energy over the planetary surface and parametrized this property with the fraction of planetary surface that re-radiates stellar flux $\beta$.
We distinguished between $\beta=0.5$, i.e. emission only from one hot hemisphere, and $\beta=1$ where the whole globe emits~\citep{Seager2005,Kaltenegger2011,Carone2014,Mendez2017}.

\begin{figure}
	\centering
\epsscale{1.2}
\plotone{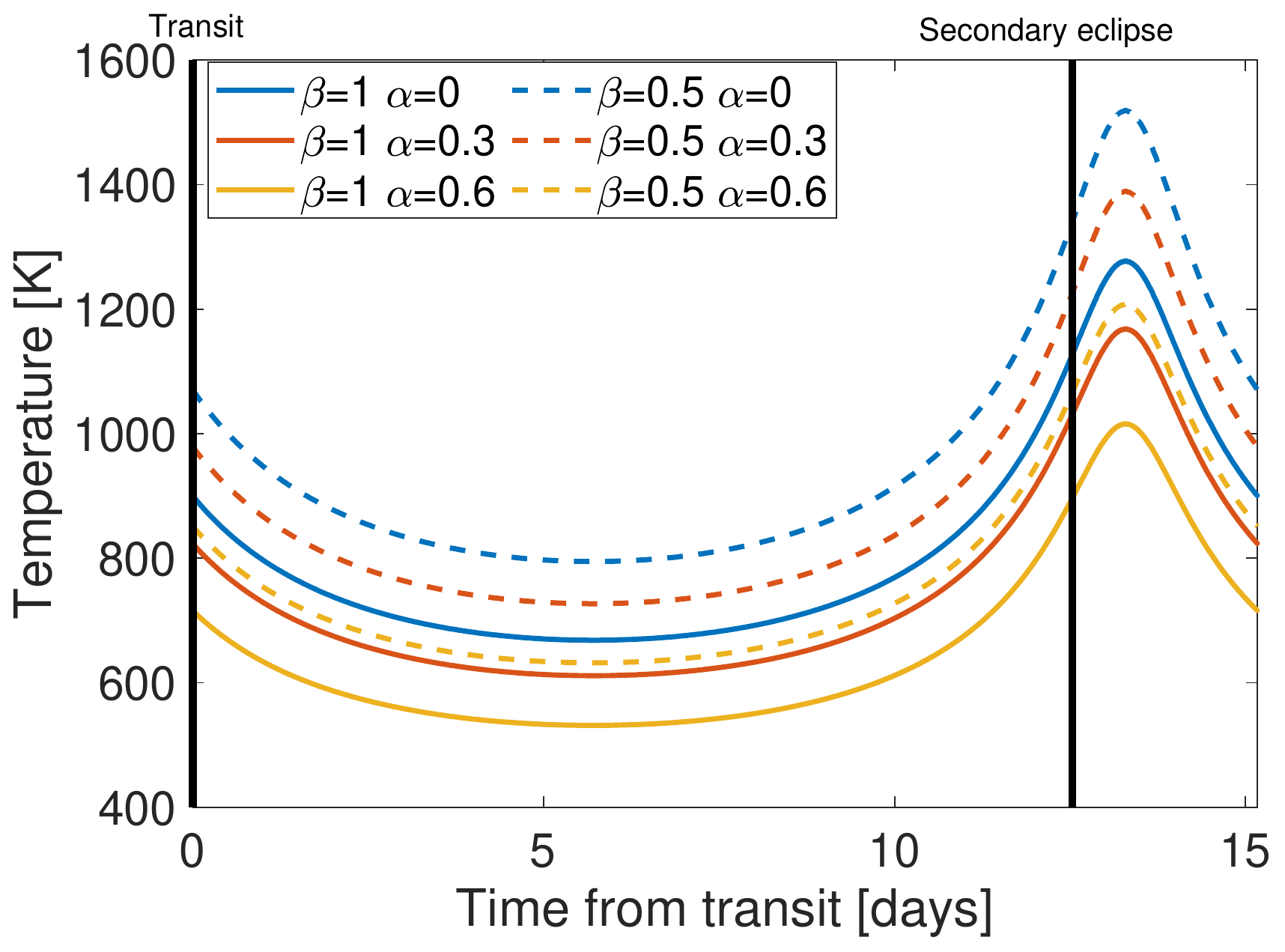}
	\caption{Evolution of the planetary equilibrium temperature in the case of instantaneous heat adjustment for different albedos $\alpha$ and re-radiation factors $\beta$.
    We assumed unity infrared emissivity $\varepsilon_\mathrm{IR}$.
    Black lines denote the time of transit and secondary eclipse, respectively.
    Due to the high eccentricity, $T_\mathrm{eq, inst}$ varies by several hundred Kelvin within one orbit.
    It stays below \SI{1000}{\kelvin} for most of the orbit.}\label{fig:Teq-theta}
\end{figure}

Planets colder than 1000~K are expected to be relatively cloudy~\citep[e.g.,][]{Stevenson2016,Parmentier2016}.
Here, we parametrized different cloudiness with albedos $\alpha=0, \,0.3$, and $0.6$ following \citet{Parmentier2016}.
With the above assumptions and in the case of instantaneous heat adjustment, we derived a range of $T_\mathrm{eq,inst} \approx \SIrange{900}{1300}{\kelvin}$ at secondary eclipse and $T_\mathrm{eq,inst} \approx \SIrange{700}{1100}{\kelvin}$ at transit (Fig.~\ref{fig:Teq-theta}).
We list $T_\mathrm{eq,inst}$ at critical times in Table~\ref{tab:T_eq} together with the values for orbitally averaged heat adjustment $T_\mathrm{eq,avg}$.
The latter is constant over one orbit and covers a temperature range of $T_\mathrm{eq,avg} \approx \SIrange{650}{975}{\kelvin}$.

\begin{figure}
\epsscale{1.2}
\plotone{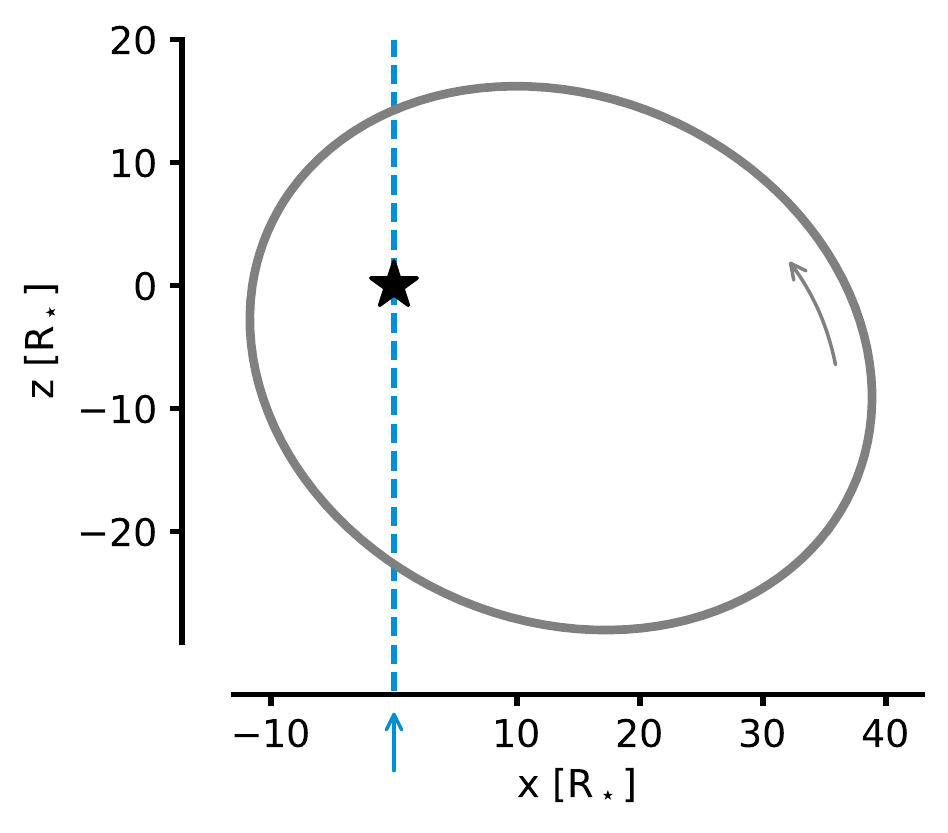}
\caption{Orbit aspect ratio and orientation. The orbit of \plname\ is plotted with stellar radii as length units.
The dashed blue line shows our line of sight with respect to the orbit.}\label{fig:systemOrientation}
\end{figure}

Due to the orientation of the orbit (compare Fig.~\ref{fig:systemOrientation}), the temperature $T_\mathrm{eq,inst}$ during transit is assumed to be about 200~K colder compared to the temperature at secondary eclipse.
In reality, however, some delay in heat adjustment based on radiative and dynamical timescales is expected.
Therefore, the temperature during a secondary eclipse, which would occur before passage of periastron, could be colder than in our estimate.
Likewise, the temperature at transit, occurring after periastron passage, would be warmer than expected~\citep[see, e.g.,][for a qualitative discussion of the thermal evolution of an exoplanet on an eccentric orbit]{Lewis2013}.
We emphasize again that our goal is to estimate to first order possible temperature ranges for \plname\, for which these simplified assumptions are sufficient.

\begin{table*}
\caption{Theoretical temperature constraints of \plname \ in the course of one orbit.}
\label{tab:T_eq}
\centering
\begin{tabular}{rc  ccc  ccc}
\hline\hline
Time & orbital distance & &$T_\mathrm{eq,inst}\,$[K] & & \multicolumn{3}{c}{$ T_\mathrm{eq,avg}\,$[K]}\\
     & [au]  & $\alpha = 0$ &  $\alpha = 0.3$ &  $\alpha = 0.6$ &  $\alpha = 0$ &  $\alpha = 0.3$ &  $\alpha = 0.6$\\
     \hline
     \multicolumn{1}{r}{$\beta=1$}& & \\

  apoastron  & 0.1900 & 668  & 611 & 531 & \multirow{4}{*}{819} & \multirow{4}{*}{749} & \multirow{4}{*}{651} \\
  transit  & 0.1047 & 900 & 823 & 716 &  &  & \\
   periastron & 0.0520 & 1277  & 1168 & 1016 & &  & \\
   secondary eclipse & 0.0668 & 1127 & 1031 & 896 & &  & \\
\hline
     \multicolumn{1}{r}{$\beta=0.5$}& & \\

  apoastron  & 0.1900 & 795  & 727 & 632 & \multirow{4}{*}{974} & \multirow{4}{*}{891} & \multirow{4}{*}{775} \\
  transit  & 0.1047 & 1070 & 979 & 851 &  &  & \\
   periastron & 0.0520 & 1519  & 1389 & 1208 & &  & \\
   secondary eclipse & 0.0668 & 1340 & 1226 & 1066 & &  & \\
   \hline
\end{tabular}

\end{table*}

\subsection{Secondary eclipses, phase curve modulations, and Rossiter-MacLaughlin effect}\label{sec:phaseCurve}

No secondary eclipses or phase curve signals are evident in the photometric time series.
We used the \texttt{starry} software package~\citep{Luger2019} to estimate the planetary phase curve based on our derived orbital parameters and a simple toy model for the planetary brightness distribution.
This model neglects any heat redistribution between the hot and cold hemispheres of the planet, resulting in a `dipole' brightness map where the hot side points to the substellar point at periastron.
In this scenario, the emission of the cold hemisphere and planetary limb darkening are negligible.
The total luminosity of the planet is then that of a half-sphere black body with radius \rpl \, and temperature $T_\mathrm{hot}$.
For the temperature of the hot hemisphere, we adopt two cases:
firstly, our estimate of the equilibrium temperature for the case of orbitally averaged heat adjustment and $\beta = 0.5$, hence $T_\mathrm{hot} = T_\mathrm{eq, avg} = \SI{974}{\kelvin}$.
Secondly, we consider the hottest temperature in Table~\ref{tab:T_eq} and assume $T_\mathrm{hot} = \SI{1519}{\kelvin}$.
With a resulting peak-to-peak phase curve amplitude of $\sim \SI{6}{\ppm}$ in the cool case and $\sim \SI{32}{\ppm}$ in the hot case, a future detection of the phase curve or secondary eclipse might be challenging despite the expected precision of the James Webb Space Telescope~\citep[JWST, ][]{Beichman2014}.

Measuring the Rossiter-McLaughlin effect~\citep[RM~effect,][]{Rossiter1924, McLaughlin1924} has proved a useful tool to measure the alignment of planetary orbits with the spin axis of host stars.
The different proposed scenarios for the formation and migration theory of warm Jupiters differ in their predicted impact on the spin-orbit alignment \citep[e.g., ][]{Triaud2018}.
A detection of the RM~effect could thus shed light on \plname's enigmatic formation history.
Analytical formulas exist to estimate the amplitude of its RV signature~\citep[e.g.,][equation 6]{Gaudi2007}, however, the large impact parameter in \plname's transit geometry would entail a large error.
Instead, we modeled the RV anomaly due to the Rossiter-McLaughlin effect with a velocity-weighted brightness map in \texttt{starry} using the median posterior values of the system's physical parameters (see Tables \ref{tab:stprops} and \ref{tab:posteriors}).

The resulting amplitude of the signal $K_\mathrm{RM} \approx \SI{10}{\meter\per\second}$, which is just in the range of current state-of-the-art spectroscopic facilities.

\todo[inline]{maybe: add comments on TTVs (i.e. that there are none evident)}

\begin{table*}
\centering
\caption{Prior parameter distributions.\\
$N(\mu,\sigma)$ stands for a normal distribution with mean $\mu$ and standard deviation $\sigma$, $U(a,b)$ stands for a uniform distribution between $a$ and $b$, and $J(a,b)$ stands for a Jeffrey's prior \rev{(that is, a log-uniform distribution)} defined between $a$ and $b$.}
\label{tab:priors}
\begin{tabular}{lccl}
\hline
\hline
\noalign{\smallskip}
Parameter name & Prior & Units & Description \\
\noalign{\smallskip}
\hline
\hline
\noalign{\smallskip}
Stellar Parameters \\
\noalign{\smallskip}
~~~$\rho_{\star}$  & $\mathcal{N}(1120,110^2) $ & kg/m$^3$ & Stellar density\\
\noalign{\smallskip}
Planetary parameters\\
\noalign{\smallskip}
~~~$P$  & $\mathcal{N}(15.16,0.2^2) $ & d & Period\\
~~~$t_{0}$  & $\mathcal{N}(2458319.17,0.2^2) $ & d & Time of transit center\\
~~~$R_\mathrm{P}/R_\star$  & $\mathcal{U}(0.0,1.5) $ & --- & Impact factor\\
~~~$b = (a/R_\star) \cos (i) $  & $\mathcal{U}(0.0,0.5) $ & --- & Planet-to-star ratio\\
~~~$K$  & $\mathcal{U}(140.0,260.0) $ & m/s & Radial velocity semi-amplitude\\
~~~$S_{1} = \sqrt{e}\sin \omega$  & $\mathcal{U}(-1,1) $ & --- & Parametrization for $e$ and $\omega$\\
~~~$S_{2} = \sqrt{e}\cos \omega$  & $\mathcal{U}(-1,1) $ & --- & Parametrization for $e$ and $\omega$\\
\noalign{\smallskip}
RV instrumental parameters\\
\noalign{\smallskip}
~~~$\mu_{\textnormal{FEROS}}$  & $\mathcal{U}(-30,30) $ & m/s & Systemic velocity for FEROS\\
~~~$\sigma_{\textnormal{FEROS}}$  & $\mathcal{J}(1.0,100.0^2) $ & ppm & Extra jitter term for FEROS\\
~~~$\textnormal{RV}_{\textnormal{linear}}$  & $\mathcal{N}(0.0,1.0^2) $ & m/s/d & Linear term for the RVs\tablenotemark{a}\\
~~~$\textnormal{RV}_{\textnormal{intercept}}$  & $\mathcal{N}(0.0,10000^2) $ & m/s & Intercept term for the RVs\tablenotemark{a}\\
\noalign{\smallskip}
Photometry instrumental parameters\\
\noalign{\smallskip}
~~~$D_{\textnormal{TESS}}$  & 1.0 (fixed)  & --- & Dilution factor for TESS\\
~~~$M_{\textnormal{TESS}}$  & $\mathcal{N}(0.0,0.1^2) $ & ppm & Relative flux offset for TESS\\
~~~$\sigma_{\textnormal{TESS}}$  & $\mathcal{J}(10^{-5},(10^{5})^2) $ & ppm & Extra jitter term for TESS\\
~~~$q_{1,\textnormal{TESS}}$  & $\mathcal{U}(0.0,1.0) $ & --- & Linear limb-darkening parametrization\\
~~~$q_{2,\textnormal{TESS}}$  & $\mathcal{U}(0.0,1.0) $ & --- & Quadratic limb-darkening parametrization\\
\noalign{\medskip}
~~~$D_{\textnormal{CHAT}}$  & 1.0 (fixed)  & --- & Dilution factor for CHAT\\
~~~$M_{\textnormal{CHAT}}$  & $\mathcal{N}(0.0,0.1^2) $ & ppm & Relative flux offset for CHAT\\
~~~$\sigma_{\textnormal{CHAT}}$  & $\mathcal{J}(10^{-5},(10^{5})^2) $ & ppm & Extra jitter term for CHAT\\
~~~$q_{1,\textnormal{CHAT}}$  & $\mathcal{U}(0.0,1.0) $ & --- & Linear limb-darkening parametrization\\
\noalign{\medskip}
~~~$D_{\textnormal{LCOGT}}$  & 1.0 (fixed)  & --- & Dilution factor for \textit{LCOGT}\\
~~~$M_{\textnormal{LCOGT}}$  & $\mathcal{N}(0.0,0.1^2) $ & ppm & Relative flux offset for \textit{LCOGT}\\
~~~$\sigma_{\textnormal{LCOGT}}$  & $\mathcal{J}(10^{-5},(10^{5})^2) $ & ppm & Extra jitter term for \textit{LCOGT}\\
~~~$q_{1,\textnormal{LCOGT}}$  & $\mathcal{U}(0.0,1.0) $ & --- & Linear limb-darkening parametrization\\
\noalign{\medskip}
\noalign{\smallskip}
Additional parameters\\
\noalign{\smallskip}
~~~$\sigma^{GP}_{\textnormal{TESS}}$  & $\mathcal{J}(10^{-8},0.0005^2)$ & --- & Amplitude of the GP component\\
~~~$\tau^{GP}_{\textnormal{TESS}}$  & $\mathcal{J}(0.0001,2^2)$ & --- & Timescale of the GP component\\
\noalign{\smallskip}
\hline
\hline
\end{tabular}
    \tablenotetext{a}{These parameters were only used to search for an additional linear RV trend and are not included in our nominal joint fit.}
\end{table*}

\begin{deluxetable}{lrc}[b]
\tablecaption{Posterior parameters}
\label{tab:posteriors}
\tablecolumns{3}
\tablehead{
\colhead{Parameter} &
\colhead{Value} \\
}
\startdata
Stellar Parameters \\
~~~$\rho_{\star} \, \mathrm{(kg/m^3)}$  & $1076^{+95}_{-93}$\\
Planetary parameters\\
~~~$P$ (d)  & $15.168865^{+0.000018}_{-0.000018}$\\
~~~$t_{0}$ (BJD UTC)  & $2458319.15055^{+0.00077}_{-0.00077}$\\
~~~$a/R_\star$  & $23.85^{+0.67}_{-0.69}$\\
~~~$b = (a/R_\star) \cos (i) $  & $0.900^{+0.017}_{-0.017}$\\
~~~$R_\mathrm{P}/R_\star$  & $0.1031^{+0.0048}_{-0.0042}$\\
~~~$K$ (m/s)  &  $191.3^{+6.4}_{-6.2}$\\
~~~$e$  & $0.575^{+0.011}_{-0.011}$\\
~~~$\omega$  & $24.1^{+2.4}_{-2.4}$\\
~~~$S_{1} = \sqrt{e}\sin \omega$  & $0.309^{+0.029}_{-0.030}$\\
~~~$S_{2} = \sqrt{e}\cos \omega$  &$0.692^{+0.014}_{-0.015}$\\
RV instrumental parameters\\
~~~$\mu_{\textnormal{FEROS}}$ (m/s)  & $14.0^{+3.4}_{-3.3}$\\
~~~$\sigma_{\textnormal{FEROS}}$ (m/s)  & $13.6^{+3.2}_{-2.6}$\\
Photometry instrumental parameters\\
~~~$M_{\textnormal{TESS}}$ (ppm)  & $0.00017^{+0.00041}_{-0.00039}$\\
~~~$\sigma_{\textnormal{TESS}}$ (ppm)  & $221^{+28}_{-31}$\\
~~~$q_{1,\textnormal{TESS}}$  & $0.33^{+0.39}_{-0.24}$\\
~~~$q_{2,\textnormal{TESS}}$  & $0.27^{+0.35}_{-0.20}$\\
~~~$\sigma^{GP}_{\textnormal{TESS}}$  & $0.00000152^{+0.00000110}_{-0.00000041}$\\
~~~$\tau^{GP}_{\textnormal{TESS}}$  & $0.40^{+0.16}_{-0.18}$\\
~~~$M_{\textnormal{CHAT}}$ (ppm)  & $0.00149^{+0.00025}_{-0.00025}$\\
~~~$\sigma_{\textnormal{CHAT}}$ (ppm)  & $1625^{+150}_{-140}$\\
~~~$q_{1,\textnormal{CHAT}}$  & $0.55^{+0.20}_{-0.24}$\\
~~~$M_{\textnormal{LCOGT}}$ (ppm)  & $-0.00038^{+0.00016}_{-0.00016}$\\
~~~$\sigma_{\textnormal{LCOGT}}$ (ppm)  & $480^{+85}_{-80}$\\
~~~$q_{1,\textnormal{LCOGT}}$  & $0.67^{+0.17}_{-0.20}$\\
Derived parameters\\
~~~i (deg) & $87.0^{+1.5}_{-1.7}$\\
~~~$R_\mathrm{P} \, [\mathrm{R_{Jup}}]$ & $1.117^{+0.054}_{-0.047}$\\
~~~$M_\mathrm{P} \, [\mathrm{M_{Jup}}]$ & $1.942^{+0.092}_{-0.091}$\\
~~~$a \, [\mathrm{au}]$ & $0.1207^{+0.0037}_{-0.0037}$\\
~~~$\rho_\mathrm{P} \, [\mathrm{kg \, m^{-3}}]$ & $1847^{+280}_{-260}$\\
~~~$T_\mathrm{eq}\, \mathrm{[K]}$\tablenotemark{a}  & \teqv \\
\hline
\hline
\enddata
\tablenotetext{a}{Time-averaged equilibrium temperature computed according to equation~16 of \citet{mendez:2017}.
We assumed zero albedo, a unity broadband thermal emissivity, and $\beta = 0.5$, i.e. only half of the planetary surface re-radiates the absorbed flux.}
\end{deluxetable}

\section{Discussion} \label{sec:discussion}

\subsection{\plname 's place in the exoplanet population}

\begin{figure*}
\gridline{\fig{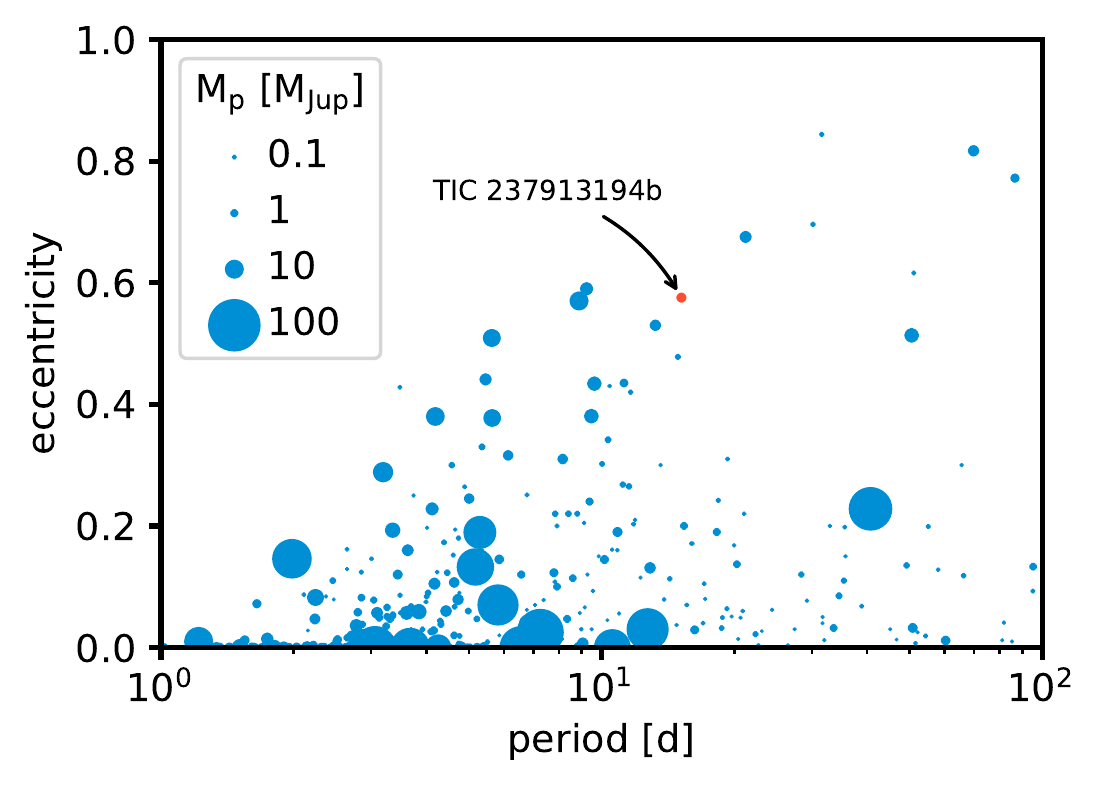}{0.492\textwidth}{}\fig{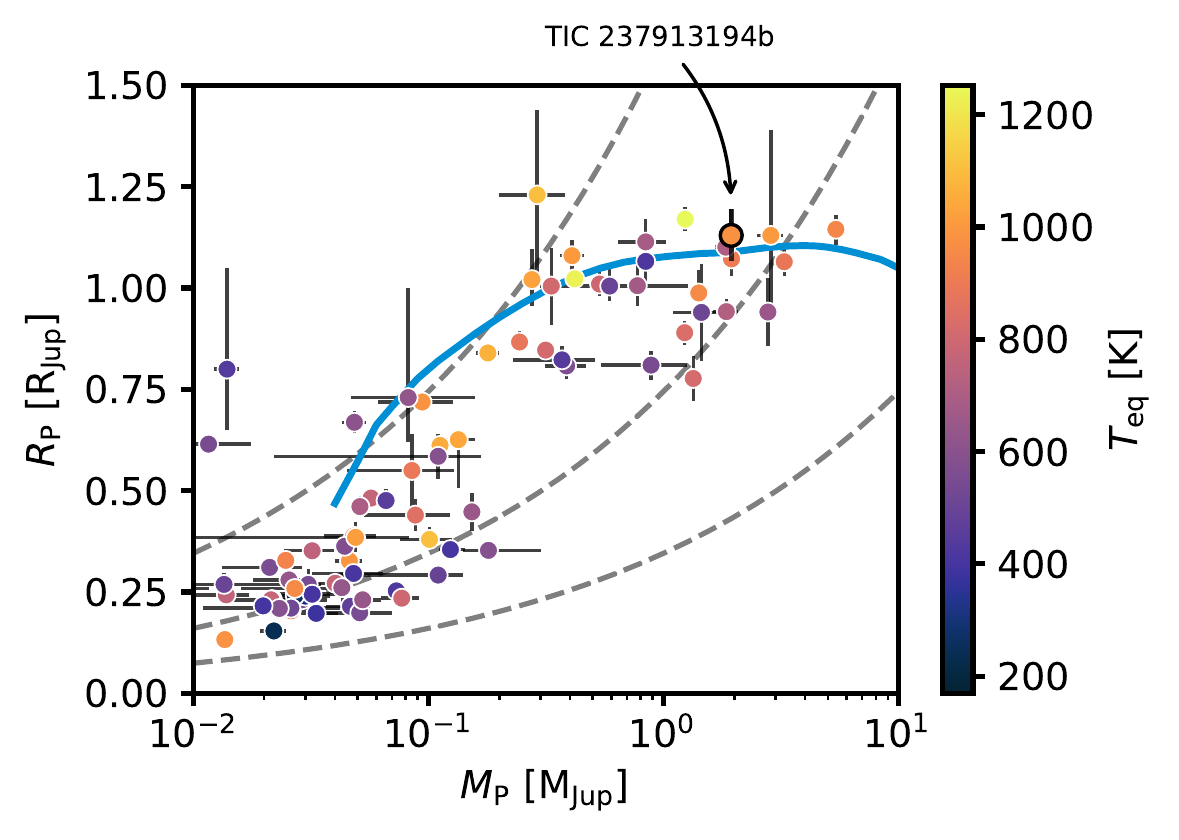}{0.52\textwidth}{}}
\caption{Comparison to other well-characterized warm Jupiters.
\textit{Left:} period-eccentricity plot of transiting exoplanets with periods of \SIrange{1}{100}{\day} and measured eccentricity from the TEPCat catalog \citep{Southworth2011}.
Marker sizes scale with planet mass.
With\rev{ $e = 0.58$, } \plname \, occupies the 98th percentile in this population and contributes to a sparse sample of planets with very high eccentricities.
~\\
\textit{Right:} mass-radius diagram of \textit{warm} ($P=\SIrange{10}{100}{\day}$) planets from the same catalog.
The color of the markers represent the equilibrium temperatures of the planets, and dashed gray lines are isodensity curves of 0.3, 3, and \SI{30}{\gram\per\centi\meter\cubed}, respectively.
The solid blue line marks the predicted mass-radius relation for giant planets with a \SI{10}{\mEarth} core~\citep{Fortney2007}.
\plname\ lies very close to this line.
The error bars for its mass are too small to be seen.
}
\label{fig:periodEcc}
\end{figure*}

In Fig.~\ref{fig:periodEcc} we compare \plname\ with well-studied transiting exoplanets~\citep{Southworth2011}\footnote{TEPCat catalog, queried on 2020-08-11.}.
The left panel shows the periods and eccentricities of these planets (blue markers); our discovery is marked in red.
We included planets with both mass and radius measurements that have constrained eccentricities (not only upper limits) and show those in the period range \SIrange{1}{100}{\day}.
Marker sizes in the plot correspond to planet masses.
Warm Jupiters with high eccentricities appear to be rare: \SI{98}{\percent} of this population have smaller eccentricities than \plname, making it one of the most eccentric planets in this period range.
It lies at the edge of a demographic feature that we discuss in the following.

On close orbits, the planet occurrence rate $\mathrm{d}n/\mathrm{d}e$ shows a rapid drop with increasing eccentricity.
With increasing period, the position of this ridge shifts to larger eccentricities.
Through this, a triangular under-density of planets with high eccentricity on very short orbits emerges (upper left corner in Fig.~\ref{fig:periodEcc}).
While exoplanet detection sensitivities are expected to have a dependency on eccentricity, the effect is too small to account for the observed dearth of planets \citep{Burke2008}.
A plausible physical explanation would be tidal circularization \citep{Adams2006, Dawson2018}.
As discussed in Sect.~\ref{sec:tidalEffects}, the strength of this mechanism is expected to scale inversely with orbital distance, which would explain the period-dependence of the distribution.
    The detection of planets close to this ridge can thus help constrain theories of tidal interaction and giant planet migration, which are crucial components for explaining planetary systems with close-in giant planets.
Our discovery of \plname \, adds to the small current sample of such planets.

In the right panel of Fig.~\ref{fig:periodEcc}, we put our planet into context of warm Jupiters with mass and radius measurements.
Here, we include only planets from TEPCat with periods of \SIrange{10}{100}{\day}, and color-code them by equilibrium temperature.
We further plot a theoretically predicted mass-radius relation for planets with a \SI{10}{\mEarth} core~\citep[][blue line]{Fortney2007}.
\plname\ is located close to this curve, which indicates that its bulk density is consistent with established structural models.

\subsection{The atmosphere of \plname}\label{sec:atmosphere}
The large eccentricity of \plname\ makes it a promising test bed to study the response of its atmosphere to external forcing~\citep[e.g., compare][]{Carone2020}.
Atmospheres at the inferred temperature ranges are susceptible to a variety of chemical disequilibrium processes such as photochemistry and chemical quenching~\citep[e.g.][]{Molaverdikhani2019a,Moses2013,Venot2012,Venot2020,Tsai2018,Kawashima2019}.
While a thorough inspection of these processes is beyond the scope of this paper, we used the physical parameters constrained here to demonstrate the feasibility of atmospheric characterization.
To this end, we used self-consistent models for cloud-free~\citep{Molaverdikhani2019} and cloudy~\citep{Molaverdikhani2020} irradiated planetary atmospheres and calculated synthetic transmission and emission spectra using \texttt{petitCODE} \citep{molliere2015model,Molliere2017}.
For this fiducial model, we assumed solar composition, zero bond albedo, instantaneous thermal equilibrium, and $\beta = 0.5$.
This resulted in an equilibrium temperature of the planet during transit and secondary eclipse of 1070~K and 1340~K, respectively (see Table~\ref{tab:T_eq}).
Using the \texttt{Pandexo} package~\citep{batalha2017pandexo}, we predicted uncertainties for JWST observations of a single transit or secondary eclipse for the three observing modes MIRI-LRS, NIRISS-SOSS, and NIRSpec-G395M.

We find relative magnitudes of the largest synthetic spectral features of $\sim$100~ppm in transmission and $\sim$1000~ppm in emission.
While some emission features are well above the predicted noise floor, the largest transmission features are on the order of the predicted uncertainties for a single transit observation. 
However, since these two techniques probe different regions of the atmosphere and at different orbital phases, a joint analysis of the transmission and emission spectra may provide important clues on atmospheric dynamics and heat redistribution.

\subsection{\plname 's large eccentricity}\label{sec:tidalEffects}
The peculiarly large eccentricity of \plname\ could be an important lead in not only understanding the dynamical origin of the system, but also planet evolution theories in general.
Possible origins of large warm Jupiter eccentricities include interaction with a massive companion through either scattering events~\citep[e.g.,][]{Rasio1996}, secular interactions~\citep[e.g.,][]{Petrovich2016,Kozai1962,Lidov1962}, or giant impacts~\citep{Frelikh2019}; planet-disk interactions~\citep[e.g.,][]{Lubow1991,Petrovich2019}; or a combination of processes.
The absence of evidence for an additional perturber that might sustainably excite \plname's eccentricity or that could be the counterpart in a recent scattering event makes it challenging to distinguish between these scenarios.

However, \plname\ is subject to tidal dissipation through secular interaction with the host star \citep[][]{Goldreich1966} and the rates of semi-major axis and eccentricity decay, given the current orbital parameters, can be estimated~\citep[][]{Yoder1981}.
A short orbit circularization timescale compared to the lifetime of the star would exclude the planet-disk interaction scenario and could provide an upper limit on the time that has passed since a hypothetical perturbation event.
In this case, we would observe the system during high-eccentricity migration and \plname\ would eventually become a hot Jupiter in a circular orbit.

Several caveats have to be considered when trying to trace the system back in time close to its primordial orbital configuration.
First, the tidal evolution of $a$ and $e$ are strongly coupled, which may result in ambiguities.
In addition, the tidal evolution strongly depends on the stellar and planetary tidal dissipation rates, typically parameterized by the dimensionless ``reduced tidal quality factors'' $Q^{\prime}_\mathrm{P}$ and $Q^{\prime}_\star$.
Here, $Q^{\prime} = 1.5\,Q / k_{2, }$, where $k_{2,}$ is the Love number of second order.
Estimates of the planetary tidal quality factor range from $Q^{\prime}_{\mathrm{P}} = 10^4$ to $Q^{\prime}_{\mathrm{P}} = 10^7$~\citep[e.g.][]{Lainey2009,Lainey2020,Hansen2012}.
The stellar tidal dissipation factor is even less well constrained but theoretical and observational works suggest $Q^{\prime}_{\star} \gtrsim 10^7$ \citep[e.g.][]{Carone2007,Hansen2012,Damiani2015}.

We studied the star-planet tides of \plname\ using the \texttt{EqTide}\footnote{\url{https://github.com/RoryBarnes/EqTide}} code \citep{Barnes2017}, which calculates the tidal evolution of two bodies based on models by \citet{Ferraz-Mello2008} and a "constant-phase-lag"~(CPL) model \citep{Goldreich1966,Cheng2014}.
For our tidal-torque test, we adopted a Stellar rotational period of 30\,d and an initial planetary rotational period of 0.5
\,d (i.e., similar to the Solar and Jupiter rotational periods).
We adopted tidal quality factors $Q^{\prime}_\mathrm{P}$ in the range of $3\times10^{4}$ - 10$^{6}$.
For the primary, we chose a fixed value of $Q^{\prime}_\mathrm{\star}$ = 10$^{8}$~\citep[e.g.,][]{Hansen2010,Penev2012,Bonomo2017}.
We tested a large set of increased initial semi-major axes and eccentricities and integrated with \texttt{EqTide}.
The results agree with the observed eccentricity and semi-major axis only for those samples that started a few percent above the observed values. %
This suggests that the orbital period of $\sim \SI{15}{\day}$ is too large for significant tidal circularization within the age of the system ($\sim$5.5 Gyr, see Table \ref{tab:stprops}) and \plname's orbit has only slightly evolved from its primordial configuration.
These results are in line with \citet{Barnes2015}, who showed that Jovians with periods longer than $\sim$8 days and a typical eccentricity of 0.3 do not experience significant orbital and eccentricity decay.
While we cannot determine the origin of the high orbital eccentricity, we conclude that the planet we observe today is not a credible progenitor of a future hot Jupiter.

\section{Conclusions}\label{sec:conclusion}

We have presented the discovery of \plname\ \rev{(TOI~2179b)}, a transiting warm Jupiter orbiting its G-type host in a very eccentric \rev{($e\approx 0.58$)} 15-day orbit.
Its transit signal was detected using \tess\ full frame images from Sectors~1 and 2.
We confirmed the planetary nature of the signal using ground-based photometry (\textit{CHAT, LCOGT}) and high-precision spectroscopy (\textit{FEROS}).
Our main results are:
\begin{itemize}
    \item a planetary mass \mpl~=~\mptess\ \mjup \, and radius \rpl~=~\rptess\ \rjup, yielding a bulk density similar to Neptune's.
    \item with \rev{$e\approx 0.58$} one of the highest eccentricities among all currently known warm giants.
    \item slow tidal evolution, prohibitive of a hot Jupiter progenitor scenario.
    \item an attractive opportunity for future observations of the planet's atmosphere, which might harbor observable chemical disequilibrium processes due to the greatly varying external forcing.
    \item good prospects for detecting the Rossiter-MacLaughlin effect, which would be a valuable contribution to the still small sample of warm Jupiters with constrained spin-orbit obliquities.
\end{itemize}

\raggedbottom %
\newpage

\acknowledgments
M.S. gratefully acknowledges insightful discussions with Robin Baeyens, Francesco Biscani, and Robert Glas.
R.B.\ acknowledges support from FONDECYT Post-doctoral Fellowship Project 3180246, and from the Millennium Institute of Astrophysics (MAS).
L.C. acknowledges support from the DFG Priority Programme SP1833 Grant CA 1795/3.
A.J.\ acknowledges support from FONDECYT project 1171208, CONICYT project Basal AFB-170002, and by the Ministry for the Economy, Development, and Tourism's Programa Iniciativa Cient\'{i}fica Milenio through grant IC\,120009, awarded to the Millennium Institute of Astrophysics (MAS).
T.H. and P.M. acknowledge support from the European Research Council under the Horizon 2020 Framework Program via the ERC Advanced Grant Origins 83 24 28.
This work was supported by the DFG Research Unit FOR2544 ``Blue Planets around Red Stars", project no. RE 2694/4-1.
\rev{This research was supported by the Deutsche Forschungsgemeinschaft through the Major Research Instrumentation Programme and Research Unit FOR2544 “Blue Planets around Red Stars” for T.H. under contract DFG He 1935/27-1 and for H.K. under contract DFG KL1469/15-1.}
This research has made use of the Exoplanet Follow-up Observation Program website, which is operated by the California Institute of Technology, under contract with the National Aeronautics and Space Administration under the Exoplanet Exploration Program.
Funding for the TESS mission is provided by NASA's Science Mission directorate.
This paper includes data collected by the \tess\ mission, which are publicly available from the Mikulski Archive for Space Telescopes (MAST).
Resources supporting this work were provided by the NASA High-End Computing (HEC) Program through the NASA Advanced Supercomputing (NAS) Division at Ames Research Center for the production of the SPOC data products.

\vspace{5mm}
\facilities{\tess, LCO, CHAT0.7m, LCOGT1.0m, MPG2.2m/FEROS}

\software{
          astropy~\citep{astropy},
          juliet~\citep{juliet},
          CERES~\citep{brahm:2017:ceres,jordan:2014},
          ZASPE~\citep{brahm:2016:zaspe,brahm:2015},
          tesseract~(Rojas, in prep.),
          TESSCut~\citep{TESSCut},
          lightkurve~\citep{lightkurve},
          radvel~\citep{fulton:2018},
          emcee~\citep{emcee:2013},
          corner.py~\citep{corner},
          MultiNest~\citep{MultiNest},
          PyMultiNest~\citep{Buchner2014},
          batman~\citep{kreidberg:2015},
          starry~\citep{Luger2019},
          celerite~\citep{Foreman-Mackey2017},
          petitCODE~\citep{molliere2015model,Molliere2017},
          EqTide~\citep{Barnes2017},
          Pingouin~\citep{Vallat2018}.
          }

\newpage

\appendix

\section{Joint fit posteriors}\label{sec:fullCorner}

\begin{figure*}
\epsscale{1.1}
\plotone{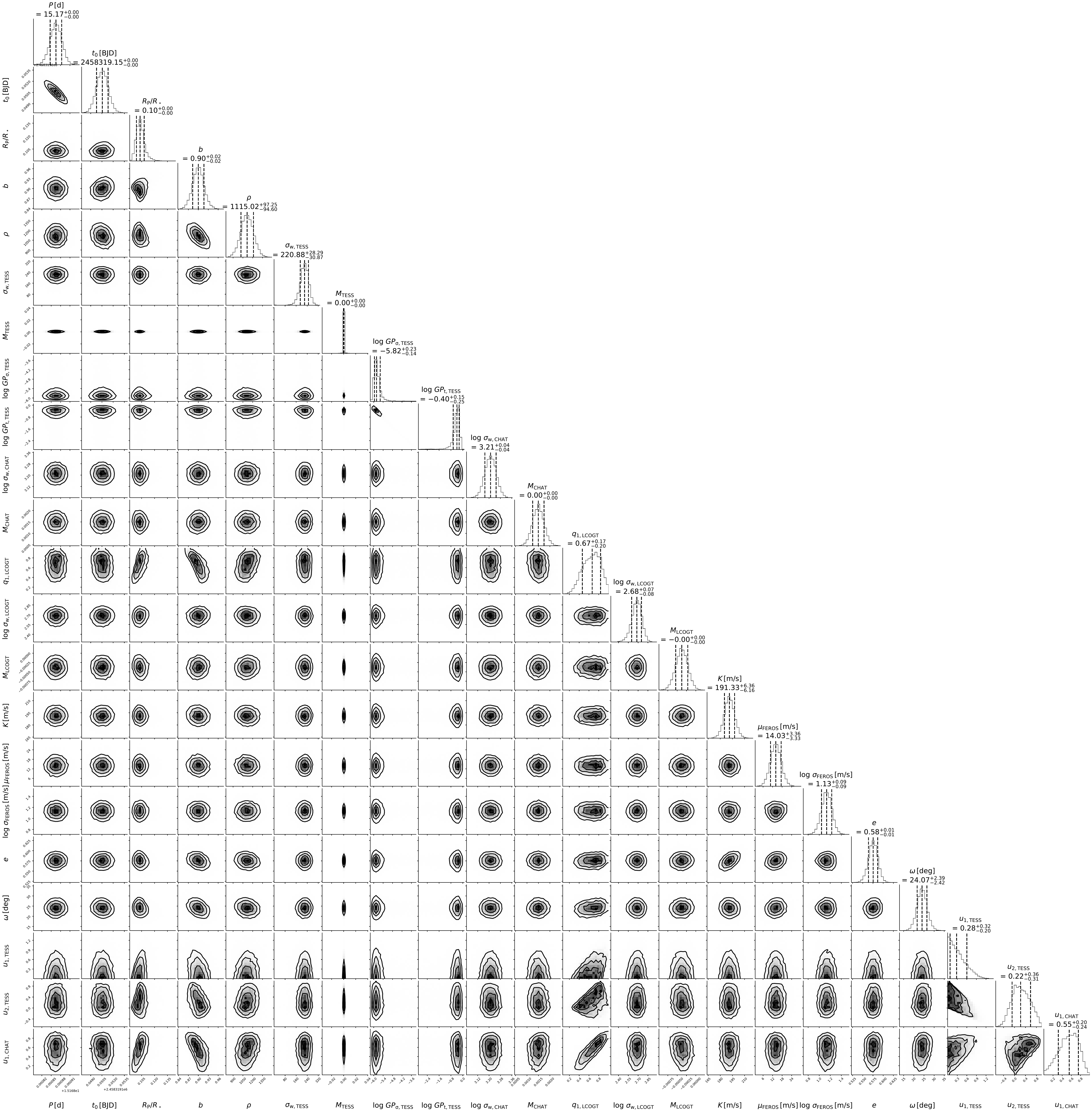}
\caption{Corner plot showing the posteriors of all parameters in our joint fit. The stated values represent 16th, 50th, and 84th percentiles, and we consider the median our `best fit'. Individual sample points are not drawn to limit file size.
} \label{fig:cornerPosteriors}
\end{figure*}
Figure \ref{fig:cornerPosteriors} shows all combinations of one- and two-dimensional projections of the posterior space from our joint fit (see Sect. \ref{sec:jointFit}).
On the diagonal, we state for each parameter the median value and its 16th and 84th percentile.

There is some residual degeneracy in the $b-\rpl/\rstar$ plane.
In previous fits that did not include \textit{LCOGT} data, the distribution extended far into the range of large impact parameters and planet-to-star ratios, marking a degenerate solution.
This effect is physically plausible: a larger planet with higher impact parameter can to some degree mimic a smaller one with lower impact parameter.
In the joint fit shown in Fig.~\ref{fig:cornerPosteriors}, this degeneracy is lifted and both $b$ and $\rpl/\rstar$ are well constrained.

\section{RV data}

\begin{table*}
\centering
\caption{FEROS radial velocities and accompanying data for \stname\ used in this paper.}
\label{tab:RVdata}
\begin{tabular}{cccccc}
\hline
\hline
\noalign{\smallskip}
BJD & RV [km/s] & $\sigma_{\rm RV}$ [km/s] & $t_\mathrm{exp}$ [s] & BIS [km/s] & $\sigma_{\rm BIS}$ [km/s]\\
\noalign{\smallskip}
\hline
\hline
\noalign{\smallskip}
2458669.798 & 29.536 & 0.009 & 1200 & -0.016 & 0.012\\
2458670.816 & 29.519 & 0.010 & 1200 & -0.014 & 0.013\\
2458672.879 & 29.533 & 0.012 & 1200 & 0.014 & 0.016\\
2458718.916 & 29.505 & 0.007 & 1200 & -0.006 & 0.011\\
2458721.768 & 29.528 & 0.007 & 1200 & -0.000 & 0.011\\
2458722.674 & 29.580 & 0.008 & 1200 & -0.001 & 0.011\\
2458723.758 & 29.595 & 0.007 & 1200 & -0.019 & 0.011\\
2458724.793 & 29.611 & 0.009 & 1200 & -0.019 & 0.012\\
2458742.740 & 29.892 & 0.009 & 1200 & -0.002 & 0.013\\
2458783.690 & 29.545 & 0.007 & 1200 & -0.033 & 0.011\\
2458785.610 & 29.646 & 0.010 & 1200 & -0.026 & 0.013\\
2458787.671 & 29.802 & 0.008 & 1200 & 0.012 & 0.012\\
2458791.610 & 29.521 & 0.009 & 1200 & -0.000 & 0.012\\
2458792.571 & 29.508 & 0.008 & 1200 & 0.007 & 0.011\\
2458800.668 & 29.614 & 0.008 & 1200 & -0.017 & 0.011\\
2458801.688 & 29.673 & 0.008 & 1200 & 0.014 & 0.011\\
2458802.619 & 29.739 & 0.007 & 1200 & -0.014 & 0.010\\
2458805.675 & 29.543 & 0.007 & 1200 & -0.001 & 0.010\\
2458847.592 & 29.701 & 0.008 & 1200 & -0.024 & 0.011\\
2458848.591 & 29.858 & 0.008 & 1200 & 0.006 & 0.011\\
2458849.592 & 29.805 & 0.008 & 1200 & 0.005 & 0.011\\
2458850.541 & 29.619 & 0.007 & 1200 & -0.010 & 0.011\\
2458911.524 & 29.560 & 0.008 & 1200 & -0.038 & 0.011\\
2458915.515 & 29.480 & 0.010 & 1200 & 0.023 & 0.013\\
2458917.514 & 29.510 & 0.010 & 1200 & 0.039 & 0.013\\
\noalign{\smallskip}
\hline
\hline
\end{tabular}
\end{table*}

\bibliography{sample63,PhD,addLit}{}
\bibliographystyle{aasjournal}

\end{document}